\documentclass[12pt,preprint]{aastex}
\usepackage{apjfonts}
\shortauthors{Wang et al.}
\shorttitle{SEDs and Morphologies of  galaxies at $z \sim 2$ }

\newcommand{\Msol}{\hbox{$M_\odot$}}
\newcommand{\Lsol}{\hbox{$L_\odot$}}
\newcommand{\uJy}{\hbox{$\mu$Jy}}
\newcommand{\um}{\hbox{$\mu$m}}

\begin{document} 

\title{CANDELS: Correlations of SEDs and Morphologies with Star-formation Status for Massive Galaxies at $z \sim 2$ }

\author{Tao Wang\altaffilmark{1,2}, Jia-Sheng Huang\altaffilmark{2}, S. M. Faber\altaffilmark{3},  Guanwen Fang\altaffilmark{4,2}, Stijn Wuyts\altaffilmark{5}, G. G. Fazio\altaffilmark{2}, Haojing Yan\altaffilmark{6}, Avishai Dekel\altaffilmark{7}, Yicheng Guo\altaffilmark{8}, Henry C. Ferguson\altaffilmark{9}, Norman Grogin\altaffilmark{9}, Jennifer M. Lotz\altaffilmark{9}, Benjamin Weiner\altaffilmark{10}, Elizabeth J. McGrath\altaffilmark{3}, Dale Kocevski\altaffilmark{3}, Nimish P. Hathi\altaffilmark{11}, Ray A. Lucas\altaffilmark{9}, A.M.Koekemoer\altaffilmark{9}, Xu Kong\altaffilmark{4}, and Qiu-Sheng Gu\altaffilmark{1}}

\altaffiltext{1}{School of Astronomy \& Space Science, Nanjing University, Nanjing 210093, China}
\email{taowang@nju.edu.cn}
\altaffiltext{2}{Harvard-Smithsonian Center for Astrophysics, 60 Garden Street, Cambridge, MA 02138, USA}
\altaffiltext{3}{University of California Observatories/Lick Observatory, University of California, Santa Cruz, CA 95064, USA}
\altaffiltext{4}{Center for Astrophysics, University of Science and Technology of China, Hefei 230026, China}
\altaffiltext{5}{Max-Planck-Institut f\"ur Extraterrestrische Physik, Giessenbachstrasse, 85748 Garching, Germany}
\altaffiltext{6}{Department of Physics and Astronomy, University of Missouri, Columbia, MO 65211, USA}
\altaffiltext{7}{Racah Institute of Physics, The Hebrew University, Jerusalem 91904, Israel}
\altaffiltext{8}{Astronomy Department, University of Massachusetts, 710 N. Pleasant Street, Amherst, MA 01003, USA}
\altaffiltext{9}{Space Telescope Science Institute, 3700 San Martin Dr., Baltimore, MD 21218, USA}
\altaffiltext{10}{Steward Observatory, University of Arizona, 933 North Cherry Avenue, Tucson, AZ 85721, USA}
\altaffiltext{11}{Carnegie Observatories, 813 Santa Barbara Street, Pasadena, CA 91101, USA}

\begin{abstract}

We  present a study on Spectral Energy Distributions, Morphologies, and star formation for an IRAC-selected extremely red object sample in the GOODS Chandra Deep Field-South. This work was enabled by new $HST$/WFC3 near-IR imaging from the CANDELS survey as well as the deepest available X-ray data from  $Chandra$ 4 Ms observations. This sample consists of 133 objects with the 3.6$\mu$m limiting magnitude of [3.6] = 21.5, and is approximately complete for galaxies with
M$_*>$10$^{11}$M$_{\odot}$ at 1.5$\leq$z$\leq$2.5. We classify this
sample into two types, quiescent and star-forming galaxies, in the observed
infrared color-color ([3.6]$-$[24] vs K$-$[3.6]) diagram. 
The further morphological study of this sample show a consistent result with the 
observed color classification. The classified quiescent galaxies are bulge
dominated and star-forming galaxies in the sample have disk or irregular
morphologies. Our observed infrared color classification is
also consistent with the rest-frame color (U$-$V vs V$-$J) classification.
We also found that quiescent and star-forming galaxies are  well
separated in the nonparametric morphology parameter (Gini vs M$_{20}$) diagram measuring their concentration and clumpiness: quiescent galaxies
have Gini coefficient higher than 0.58 and star forming galaxies have Gini
coefficient lower that 0.58. We argue that the star formation quenching
process must lead to or be accompanied by the increasing galaxy concentration. One prominent morphological feature of this sample is that
disks are commonly seen in this massive galaxy sample at 1.5$\leq$z$\leq$2.5: 30\% of quiescent galaxies and 70\% of star forming galaxies with
M$_*>$10$^{11}$M$_{\odot}$ have disks in their rest-frame optical morphologies. The prevalence of these extended, relatively undisturbed disks challenges 
the merging scenario as the main mode of massive galaxy formation.

\end{abstract}
\keywords{galaxies: evolution --- galaxies: formation --- galaxies: high-redshift --- galaxies: structure}
\clearpage

\section{Introduction}
Understanding when and how the most massive galaxies formed remains one of the most outstanding problems in galaxy formation. 
In the present-day universe, most massive galaxies ($M \ga 10^{11} \Msol$) are  early-type galaxies \citep{Baldry:2004}, which are primarily composed 
of old stellar populations with no or little star formation. Spectroscopic studies of large local samples reveal that the bulk of 
stellar mass in massive early-type galaxies was formed at $z \ga 1.5$ within a short timescale \citep[and refs. therein]{Thomas:2005,Renzini:2006,Jimenez:2007}. 
At intermediate redshifts, several surveys reveal that the precursors of present-day massive  early-type 
galaxies have already turned passive by $z \sim 0.8$, but may continue to assemble their stellar mass later through dry 
mergers \citep{Bell:2004,Yamada:2005,Bundy:2006,Cimatti:2006,Borch:2006,Faber:2007,Brown:2007}. The lack of significant star 
formation in these galaxies since $z = 0.8$ implies that their stellar mass formed at higher redshifts. Recently, both 
observational and theoretical results suggest that the critical epoch for massive galaxy evolution and mass assembly is at $1 
\la z \la 3$ \citep{Fontana:2004,Glazebrook:2004,Nagamine:2005,Kong:2006,Papovich:2006,Abraham:2007,Arnouts:2007,Ilbert:2010}. Thus, a complete census of a well-defined massive galaxy sample at this redshift range is crucial to understanding the formation and evolution of these objects. 

%Because of the low efficiency of obtaining optical or near-infrared spectra at high-redshifts, large spectroscopic surveys become progressively less efficient at 
Significant progress has been made in detecting and identifying massive galaxies at $1 \la z \la 3$ using large-format 
near-infrared (NIR) array cameras on 4$-$10 m class telescopes. 
Recent wide-field NIR surveys reveal several "new" galaxy populations, which includes Extremely Red Objects (EROs; $R - K > 5$, 
\citealt{Elston:1988}), IRAC-selected EROs ( $z_{850} - [3.6]$>$ 3.25$\footnote{$z_{850}$ is the F850LP band magnitude.}, 
\citealt{Yanh:2004}), BzKs (sBzK:$(z - K) - (B - z) > -0.2$, pBzK: $z - K > 2.5$; \citealt{Daddi:2004}), and Distant Red Galaxies 
(DRGs, $J - K > 2.3$, \citealt{Franx:2003}). These galaxy populations are mainly selected on the basis of their red rest-frame 
optical colors, and have typical stellar masses in excess of $10^{11}\Msol$ down to current NIR limits. Moreover, various studies 
show that they dominate the high-mass regime of the $z \sim 2$ stellar mass function (SMF). For instance, \cite{Conselice:2008a} 
suggested that EROs contribute 75\% of all $M_{*} > 10^{11}$$\Msol$ galaxies at $z \sim 1.5$ down to $K_{Vega}$ = 19.7, while 
\cite{van_Dokkum:2006} found a similar fraction for DRGs in the range of $2 < z < 3$. 
This makes red color selection an efficient and effective way of studying massive high-redshift galaxies.
%%Among novel techniques for isolating high-redshift galaxiesIn contrast to galaxies selected based on their rest-frame UV colors,e.g., Lyman-break galaxies, BM and BX galaxies \citep{Steidel:2003,Adelberger:2004,Steidel:2004} which 
%Thus, in contrast to UV-selected galaxies which are more sensitive to "normal" and less massive star-forming galaxies, have typical stellar masses Given the high efficiency of isolating massive galaxies compared to UV-selected galaxies
%Thus red galaxies provide an efficient method. advantage that contaminates from low-z objects
%What we have known about 

% compared to LBGs?

Utilizing these novel techniques for identifying high-redshift massive galaxies, a great deal of work has been carried out to 
explore their nature. These galaxies turn to be very different from their local counterparts. First, star formation activities were more intense at higher redshifts. At $z \sim 2$, more than half of the massive galaxies are rapidly star-forming galaxies  with $\ga$ 100$\Msol$yr$^{-1}$ while the remaining are quiescent galaxies with no or little ongoing star-formation 
\citep{Papovich:2006,Daddi:2007,Cassata:2008,Kriek:2008a}. In contrast, at $z \sim 0$ most of the galaxies with similar mass are quiescent galaxies. Second, while the cosmic mass density in massive star-forming galaxies is flat or decreases from $z \sim 2$ to the present day, the mass density in massive quiescent galaxies increases by a factor of $\sim$ 10 over the same time period. Such dramatic evolution can not be fully explained by  modest mass growth of individual quiescent galaxies. Transformation of the star-forming galaxies into quiescent galaxies is indispensable to match this density evolution of quiescent galaxies, as well as that of star-forming galaxies themselves \citep[see, e.g.,][]{Brammer:2011}. Understanding this transformation is now one of the central questions in studying galaxy evolution.

%SFR , among the most fundamental quantity, stellar mass assembly --current while morphology assembly history.
%AGN contamination in the color-Morphology. photometric errors 

%Spectroscopic follow-up of massive galaxies are extremely difficult due to their optical faintness (typically $R > 25$) and the low efficiency of near-IR spectroscopy. largely rely on multi-wavelength photometric methods. confirmed by spectra. 
%Successful spectroscopic studies, though limited to small samples, show consistent results with photometric 
%\citep[SFGs,][]{Cassata:2008,Kriek:2008a}. reveal their diverse nature: roughly half of them are quiescent with no or little ongoing star-formation while the other half are star-forming galaxies. This indicates that many massive SFGs at $z \sim 2$ eventually transformed into quiescents 
%after quenching of star-formation.
To study the physical mechanisms responsible for this transformation, it is essential to classify galaxies into different types.
Though a clear bimodal color distribution between quiescent and star-forming galaxies persists to at least $z \approx 1$, the bimodality becomes less 
distinct at the high-mass end \citep[e.g.,][]{Strateva:2001,Blanton:2003,Baldry:2004,Bell:2004,Faber:2007}. This may be largely 
attributed to an increase in dust content in the massive star-forming galaxies which reddens their colors \citep{Baldry:2004,Baldry:2006,Driver:2007}. The dust reddening turns to be more significant at higher redshifts due to increasing importance of obscured star-formation. At $z \sim 2$ most massive star-forming galaxies have very red optical-to-infrared colors. Indeed, these star-forming galaxies have such red colors that they are prematurely on the red sequence \citep{Brammer:2009}. %A number of other methods, were proposed to separate these red, dusty galaxies from the red, quiescent galaxies. 
% thus the bimodality in the color-mass diagram can no longer be seen.
A number of other methods were proposed to separate the two populations, using for example, observed colors $I - J$ 
vs. $J - K$ \citep{Pozzetti:2000}, rest-frame colors  $U - V$ vs. $V - J$ \citep{Wuyts:2007,Williams:2009}, mid-infrared colors \citep{Papovich:2006,Fontana:2009}, or directly using 
output spectral types and specific star-formation rates (SSFR\footnote{ SFR per stellar mass, SSFR = SFR/M$_{*}$}) from the 
SED fitting process \citep{Arnouts:2007,Damen:2009,Ilbert:2010,Guo:2011}. Each of these techniques has its advantages and disadvantages, and they can give different classification results for a significant number of galaxies. Resolving these discrepancies is important for us to have a clean separation of quiescent and star-forming populations. Moreover, the fraction of AGNs among massive galaxies at $z \sim 2$ is high, up to $\sim$ 30\% \citep[see, e.g.,][]{Messias:2010}. AGN activity can also contribute to the infrared emission, which is used to identify star-forming galaxies in most of the classification methods. Thus a comprehensive identification of AGN population is necessary for a better understanding of galaxy classification.
%Understanding whether these criteria select exactly the same population is important Each of these techniques has advantages and disadvantages, understanding Moreover, the fraction of AGNs among massive galaxies at $\sim 2$ is high, up to $\sim$ 30\% \citep[see, e.g.,][]{Messias:2010}. Thus identification of AGNs is also necessary in galaxy classification as well as in understanding the roles of AGNs in galaxy transformation.

%the importance of identifying AGN

In addition to the star-formation status, morphologies provide important constrains on the galaxy assembly history and are 
crucial to distinguish among models of galaxy formation \citep{Roberts:1994}. Studies of morphologies for high-redshift galaxies 
have greatly advanced in the last decade \citep{Papovich:2005,Lotz:2006,Toft:2007,Buitrago:2008,Cameron:2011,Forster:2011b,Kartaltepe:2011}. 
Several recent work present evidence that galaxy morphologies are correlated with star-forming status for galaxies at $z \sim 2$: quiescent galaxies tend to be spheroidal and compact while star-forming galaxies are mostly disks or irregular/mergers with larger sizes \citep{Daddi:2005,Toft:2007,van_dokkum:2008,Buitrago:2008,Kriek:2009b,Damjanov:2009,Szomoru:2010,Cassata:2011,Weinzirl:2011,Wuyts:2011b,Bell:2011}.
However, this correlation at $z \sim 2$ shows large scatter and is far from clear. A significant number of systems are classified as having the same star-formation status yet with different morphological types \citep[see, e.g.,][]{Conselice:2011b}. Based on the new $HST$/WFC3 NIR imaging, \cite{van_der_Wel:2011} shows that $\sim$ 65\% of quiescent galaxies are actually disk-dominated \citep[also see,][]{Cameron:2011}, though their sample is small . Moreover, many of these work characterized galaxy  structures with the S{\'e}rsic index obtained by fitting a S{\'e}rsic model to the surface brightness data, i.e., a model-dependent approach. Yet there are concerns whether the profile of a high-redshift galaxy can be well fitted with a single s{\'e}rsic profile \citep{Szomoru:2010}. Thus other methods, including both visual classification and model-independent measurement of galaxy morphology, are needed to probe the relationship between galaxy morphology and their physical properties. Nevertheless, this requires high-resolution and high signal-to-nosie NIR (rest-frame optical) imaging for a statistical sample of $z \sim 2$ galaxies, which now becomes available thanks to the Cosmic Assembly Near-infrared Deep Extragalactic Legacy Survey \citep [CANDELS; PI: Sandra M. Faber and Henry C. Ferguson, cf, ][]{Grogin:2011, Koekemoer:2011}. %, such as star-forming status and masses.      

In this paper, we present a study on star-formation activity, morphologies, and rest-frame colors for a sample of massive galaxies at $z \sim 2$ selected in the GOODS-South. This work was enabled by the new $HST$/WFC3 near-IR imaging from the CANDELS as well as the deepest available X-ray data from  $Chandra$ 4 Ms observations. These new data, supplemented with the available rich data set in this field, allows us to have a better separation between quiescent galaxies, star-forming galaxies, and AGNs.
We use a sample of bright IRAC-selected Extremely Red Objects \citep{Yanh:2004}, which effectively reject low-redshift contaminates. \cite{Conselice:2011a} argue that the IERO selection is the most efficient among other methods (EROs, $BzK$s, and DRGs) for identifying massive galaxy samples at $1.5 < z < 3$.

The structure of this paper is as follows. The sample selection is described in section 2. We discuss the classification between quiescent and star-forming galaxies based on their star-formation status in section 3. Both visual classification and nonparametric approaches are performed on their $HST$/WFC3 F160W morphologies in section 4. After separating quiescent and star-forming galaxies, we derive their redshift and stellar mass in section 5. We examine the joint distribution of star formation activity, morphology and rest-frame colors in section 6. A summary of our main results is given in section 7. 
Throughout the paper, we assume $\Omega_\Lambda = 0.7$, $\Omega_M = 0.3$, and $H_{0}$ = 70 km s$^{-1}$ Mpc$^{-1}$. All magnitudes are in the AB system  unless specified otherwise, and the notation  "$[3.6]$" means the AB magnitude at wavelength 3.6 $\micron$.

\section{ Data Sets and IERO Selection} 
The GOODS Chandra Deep Field-South (GOODS-S, \citealt{Giavalisco:2004}) is among the best studied deep fields, with a rich multi-wavelength data set. The ultradeep IRAC imaging of GOODS-S covers the whole $10' \times 15'$ region (Dickinson et al., in preparation) and has  an exposure time of 23 hrs per pointing, yielding a 5$\sigma$ limiting magnitude of [3.6] $\sim$ 25.6. Zheng et al. (2012, in preparation) reprocessed all GTO and legacy IRAC survey data in five CANDELS fields including the GOODS-S to produce a homogeneous catalogue. We adopted their 3.6 $\um$ selected sample as our primary catalogue. 

\cite{Giavalisco:2004} performed very deep ACS imaging including F435W, F606W, F775W, and F850LP bands in the GOODS-S, with a 5$\sigma$ limiting magnitude of $z_{850} \sim 27.3$. \cite{Retzlaff:2010} also carried out deep VLT/ISAAC NIR imaging in this field, with  limiting magnitudes $J$ = 25.0, $H$ = 24.5, and $K_{s}$ = 24.4 (5$\sigma$). The deep NIR photometry is essential for deriving accurate photometric redshifts and stellar masses for red galaxies at $z \sim 2$ \citep{Wuyts:2007}.

MIPS 24 $\um$ \citep{Rieke:2004} is very sensitive to dust emission from galaxies in the redshift range $0 < z < 3$ \citep{Huang:2005,Huang:2009,Yanl:2005,Webb:2006,Papovich:2007}, and is thus extremely useful to identify the dusty star-forming population in red galaxies. The  24 $\um$ imaging in the GOODS-S is the deepest ever obtained with an exposure time of 10 hour per pointing. The 3$\sigma$ point-source sensitivity limit is $\sim$10$\uJy$, permitting detection of  luminous infrared galaxies (LIRGs) with $L_{IR} \sim 2 \times 10^{11} \Lsol$ out to $z \sim 2.5$. On the other hand, the 24 $\um$ emission from a galaxy can  also be powered by an AGN, which can be identified by deep X-ray observations. 
The recent  4 Ms $Chandra$ imaging in the GOODS-S, the deepest $Chandra$ survey ever obtained, reaches on-axis flux limits of $\approx 3.2 \times 10^{-17}$, $9.1 \times 10^{-18}$, and 5.5 $\times$10$^{-17}$ erg cm$^{-2}$ s$^{-1}$ for the full ($0.5-8$ keV), soft ($0.5-2$ keV), and hard bands ($2-8$ keV), respectively \citep{Xue:2011}. These deep X-ray data permit detection of even highly obscured AGNs.

A major breakthrough in deep imaging of GOODS-S comes from two recent \textit{HST}/WFC3 NIR surveys: the WFC3 Early Release Science (ERS, \citealt{Windhorst:2011}) program and the first phase of the CANDELS survey. The ERS mosaic covers the northern $10' \times 4'$ region of GOODS-S while the CANDELS dataset  covers the  southern $10' \times 12'$ region. Both surveys reach 5$\sigma$ limiting magnitude of  $\ga$ 27.0 in F160W ($H$-band), providing high-quality structural information (concentrations, Gini/M$_{20}$) for galaxies as faint as $H$ = 24.6, which is equivalent to 2 $\times$ 10$^{10} \Msol$ at $z \sim 2$ \citep{Grogin:2011}.
 
\begin{figure}
\center{\includegraphics{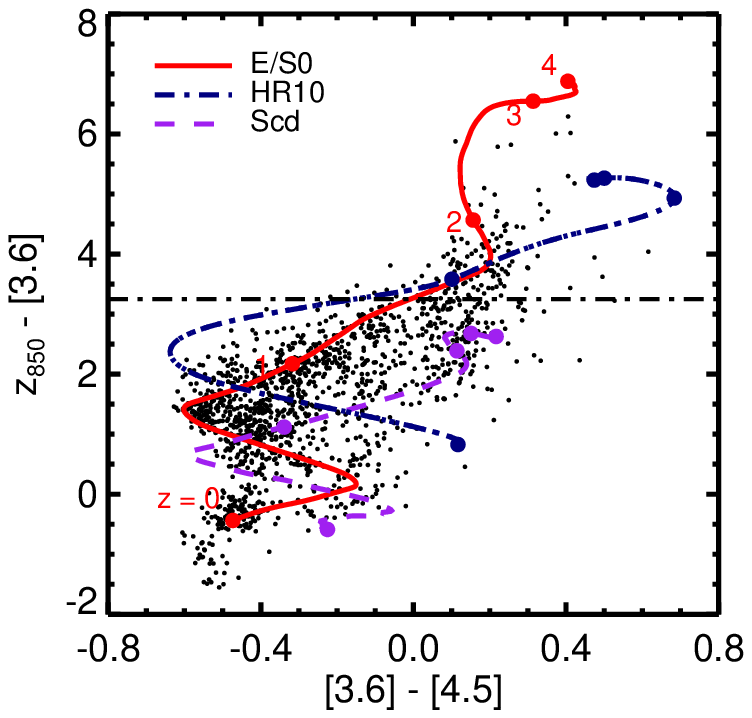}}%{paper1_z36_ch12.eps}}
\noindent
\caption{
Observed-frame $z_{850}$ $-$ [3.6] vs. [3.6] $-$ [4.5] color-color diagram for all IRAC-selected ([3.6] $<$ 21.5) galaxies in the GOODS-S. 
Observed-frame color tracks for nonevolving E/S0, Scd SEDs \citep[CWW E,][]{Coleman:1980}, and a dusty star-forming galaxy \citep[HR10,][]{Dey:1999,Stern:2006} SED at $0 < z < 4$ are plotted using color-coded lines. The IERO color definition, $z_{850}-$[3.6] $>$ 3.25, is marked with a black dashed line. The red $z_{850}-[3.6]$ color can be produced by both old and dusty galaxies at $z \ga 1$. \label{ch12_z36_sed}}
\end{figure}

The sample is selected within the central 138 arcmin$^{2}$ area in the GOODS-S where deep ACS, WFC3, ISAAC, and IRAC data are all available. We set criteria 
\begin{equation}
      [z_{850}] - [3.6] > 3.25~~~~~~and~~~~~~[3.6] < 21.5
\end{equation}
\label{eq:iero} 
to select a bright IERO sample. Figure~\ref{ch12_z36_sed} shows that the color criterion $[z_{850}] - [3.6] > 3.25$ mainly selects quiescent galaxies and dusty star-forming galaxies at $z \ga 1$ \citep{Yanh:2008}. Their red colors are due to either the Balmer/4000 \textup{\AA} break shifting beyond F850LP or their steep slope of the rest-frame UV spectra. Young and blue galaxies can not reach this threshold even at higher redshifts. Thus simple $z_{850} - [3.6]$ cut are very effective in selecting galaxy samples at high-redshift. Moreover, the bright magnitude cut of $[3.6] < 21.5$ ensures that all galaxies in this sample have high-quality SEDs as well as reliable morphological measurements. Our final IERO sample includes 133 galaxies.
The median $z_{850}$ band magnitude in our sample is $z_{850} \sim$ 24.6, and 80\% of them have at least a 3$\sigma$ detection in all the ACS and IRAC bands. For comparison, the IERO sample in \cite{Yanh:2004} is much fainter with a median $z_{850} \sim 27$. 
We then matched these sources to the FIREWORKS catalog \citep{Wuyts:2008} and obtained multi-wavelength photometry from ground-based $U$ band to \textit{Spitzer}/MIPS 24 $\um$.
29\% of the sample, or 38 sources, have an X-ray counterpart in the 4 Ms $Chandra$ source catalog within a search radius of 1.5$''$. 
%All of these X-ray detected IEROs are classified as AGNs according to their X-ray luminosities \citep{Xue:2011}, indicating that the AGN fraction in this sample reaches at least $\sim$ 30\%. 
% 84\% of them, 32 sources, are moderate-luminosity AGNs with $L_{X} \sim 10^{42-44}$ erg s$^{-1}$, and the remaining six sources are QSOs with $L_{X} > 10^{44}$ erg s$^{-1}$.

Figure~\ref{ch12_z36_sed} also shows the distribution of their [3.6] $-$ [4.5] colors for this sample, which can be used to constrain their redshifts. Most galaxies in the sample have [3.6] $-$ [4.5] $>$ 0, indicating that they are mainly at $z$ $\ga$ 1.5 when the 1.6 $\um$ stellar emission bump begins to move into the IRAC 4.5 $\um$ band \citep{Sawicki:2002,Huang:2004,Papovich:2008,Huang:2009}. Their photometric redshifts are presented in section \ref{photoz}.

\section {Separating IEROs into Quiescent and Star-forming Populations based on [3.6] $-$ [24] color}
\label{class}
We have shown that the IEROs tend to be either quiescent galaxies (quiescents) or dusty star-forming galaxies (dSFGs) at $z > 1.5$. However, it is difficult to separate them solely based on observed UV-to-NIR SEDs. Instead, the mid-infrared, e.g., the MIPS 24 $\um$, traces star-forming activity across a wide range of redshifts, and can be used to separate these two populations \citep{Papovich:2006,Fontana:2009,Fang:2009}. In particular, in redshift $1.5 < z < 2.5$ the 7.7 $\um$ PAH emission shifts into the observed 24 $\um$, allowing a universal conversion of 24 $\um$ flux into star-formation rate (SFR,\citealt{Papovich:2007,Rieke:2009,Elbaz:2011}). On the other hand, the IRAC 3.6 $\um$ in this redshift range probes rest-frame NIR, which can be treated as a good proxy for stellar mass \citep{Bell:2001,Cole:2001}. 
%Deriving SSFR for galaxies requires knowledge of their redshifts, stellar populations and SFR, for which explicit determinations have large uncertainties for extremely red galaxies. 
%Here, since we are primarily attempting to make a separation of the two types of red galaxies, 
%We propose to use the [3.6] $-$ [24] color as a proxy for specific star-formation rates (SSFR) for galaxies in our sample. This sample mainly includes massive galaxies at $1.5 < z < 2.5$ (Figure~\ref{ch12_z36_sed}, \cite[also see,][]{Yanh:2004}). In this redshift range, the 7.7 $\um$ PAH emission 
%shifts into the observed MIPS 24 $\um$ while the IRAC 3.6 $\um$ probes rest-frame NIR.  
Thus we argue that the [3.6] $-$ [24] color provides a good indicator of specific star-formation rates (SSFR) for galaxies in this sample. 
%It is also possible for quiescent galaxies to have substantial 24 $\um$ emission if they host an AGN. Deep X-ray observations are the most direct way to identify AGNs. Utilizing the 4 Ms Chandra Substantial 24 $\um$ flux can also result from AGN activity. 

\begin{figure*}
\center{\includegraphics[width=0.9\textwidth]{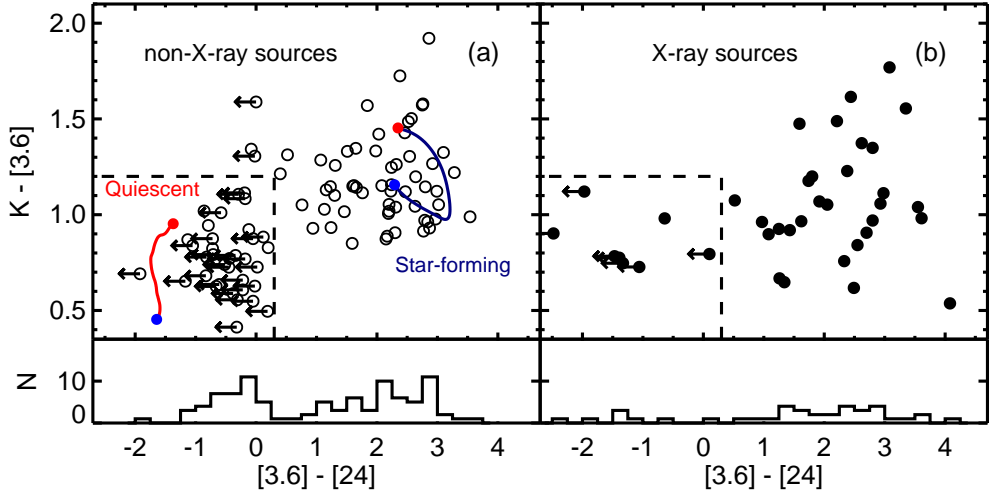}}%{paper1_k36_ch15_all.eps}}
   \caption{The [3.6] $-$ [24] vs $K - [3.6]$ color-color diagram for galaxies in our sample, which is divided into non-X-ray (left panel) and X-ray sources (right panel). Upper limits correspond to galaxies fainter than 11 $\uJy$ ($3 \sigma$ limiting flux density) in the 24 $\um$ image. The color tracks for the CWW E/S0 and HR 10 SEDs at $1.5 < z < 3$ are shown in red and blue lines. The blue and red filled circle indicate the locus of their colors at $z = 1.5$ and $z = 3$, respectively. The [3.6] - [24] color provides an estimate of SSFR while the $K - [3.6]$ color is sensitive to dust extinction. The [3.6] $-$ [24] color distribution is clearly bimodal. All the 24 $\um$-undetected IEROs are at [3.6] $-$ [24] $<$ 0.3, most of which also have bluer $K - [3.6]$ colors. We propose a preliminary color criteria of [3.6] $-$ [24] = 0.3 and $K - [3.6] = 1.2$ to separate quiescent and star-forming galaxies.  The central AGN can also have significant contribution to their 24 $\um$ emission, and thus this criterion may not apply to some X-ray IEROs.
 \label{k36_ch15}}
  \end{figure*}

We plot K $-$ [3.6] vs. [3.6] $-$ [24] color-color diagram for X-ray undetected and X-ray detected IEROs in Figure~\ref{k36_ch15}a and Figure~\ref{k36_ch15}b, respectively. 
The  IEROs show a clearly bimodal distribution in the [3.6] $-$ [24] color. The 3$\sigma$ limiting flux density at 24 $\um $ in the GOODS-S is $\sim 11 \uJy$, equivalent to a SFR of $\sim $ 10 $\Msol$ yr$^{-1}$ at $z \sim 2$ \citep{Rieke:2009}. IEROs have a typical stellar mass of $\sim 10^{11} \Msol$ \citep[see,e.g,][]{Yanh:2004}, thus this 24 $\um$ detection limit is equivalent to a SSFR $\sim 10^{-10}$ yr$^{-1}$. All the 24 $\um$-undetected galaxies are in the $[3.6] - [24] < 0.3$ region in Figure~\ref{k36_ch15}, and we propose an empirical color criterion, [3.6] $-$ [24] = 0.3, to classify this sample into quiescents and dSFGs. 
%, we estimate that the criterion [3.6] $-$ [24] = 0.3 is  equivalent to a SSFR $\sim 10^{-10}$ yr$^{-1}$ for galaxies in our sample
This criterion is consistent with that used in \cite{Damen:2009}\footnote{\cite{Damen:2009} adopted $SSFR < 1/(3 \times t_{H}$) yr$^{-1}$ to select quiescents, where $t_{H}$ is the age of the universe at a given redshift. This criterion is used to characterize the low-SSFR peak of massive galaxies, which increases with redshift. At $z = 2$ this yields $SSFR$ $\la 10^{-10}$ yr$^{-1}$.} We also verified that by changing the [3.6] $-$ [24] color criterion within a scatter of 0.2, the numbers in the quiescents and dSFGs population changed at most $\sim$10\%, and our main conclusions drawn in this paper remain valid.

%Several dSFGs in our sample may have very weak or absent 24 $\um$. 

In several cases, 24 $\um$ emission may be very weak or absent even for dSFGs. At $z \sim$1.4, the silicate absorption at 9.7$\um$
shifts in the MIPS 24 $\um$ band. Thus SFGs with a strong silicate absorption feature may have no 24 $\um$ detection at this redshift \citep{Magdis:2011}. We thus propose another  K $-$ [3.6] color to At $z \sim 2$, K $-$ [3.6] probes rest-frame $R - J$, which is sensitive to dust extinction. DSFGs have redder K $-$ [3.6] colors compared to quiescents at similar redshifts, as shown by the color tracks of a quiescent and a dSFG SED template in Figure~\ref{k36_ch15}a. 
Most objects in our sample with $[3.6] - [24] < 0.3$ have blue K $-$ [3.6] , but a few of them have very red K $-$ [3.6] (see galaxies to left of dashed line in Figure~\ref{k36_ch15}(a)).  Photometric redshifts
for these sources confirm that they are at $z \sim1.4$ or at very high redshift, $z \sim 3$, where 24$\um$ is not sensitive to star formation.  We argue that these sources are really dSFGs. We therefore propose the observable 
color criteria for identification of quiescents in the sample as:

\begin{equation}
	\label{eq:IR_color}
      [3.6] - [24] < 0.3 ~~and~~  K - [3.6] < 1.2. ~~~~~~
\end{equation}
The remaining  objects are denoted as dSFGs. We note that though the adopted $K - [3.6]$ color criterion is empirical, the number of these 24 $\um -$faint dSFGs is very small and have redshifts beyond our main interests.

%This [3.6] $-$ [24] criterion may not apply to quiescent galaxies with AGNs, which can also have strong 24 $\um$ emission. For example, \cite{Gu:2007} showed an elliptical galaxy, NGC 315,that has MIR and even FIR emission powered by a central AGN. The 4 Ms $Chandra$ imaging can detect an AGN with $L_{X} \sim 10^{42}$ erg s$^{-1}$ at $z \sim 2$.
%38/133 galaxies, 29\% of the sample, are detected in the 4 Ms $Chandra$ imaging, all of which are classified as AGNs according to their X-ray luminosities \citep{Xue:2011}. The [3.6] $-$ [24] and K $-$ [3.6] colors for these X-ray sources are shown in  Figure~\ref{k36_ch15}b. Several AGNs even show very blue K $-$ [3.6] contrary to their red [3.6] $-$ [24]. Further classification of these X-ray IEROs will be performed using their morphologies and rest-frame colors.
  
%This criterion is only for IEROs with no X-ray detections. We will have further classification for the X-ray sources using their morphologies and rest-frame optical-NIR colors.

Based on Eq~\ref{eq:IR_color}, a total of 90 galaxies, 68\% of the whole sample, are classified as dSFGs; 43 galaxies, 32\% of the whole sample, are quiescents. 
The median [3.6] $-$ [24] colors for the two populations are  $\sim$ 2.2 and  $-$0.5, respectively, indicating that their SSFRs are significantly different. In fact, those galaxies with [3.6] $-$ [24] $>$ 0.3 have a median 24 $\um$ flux density of  $\sim$ 125 $\uJy$, implying that a significant fraction of them are ultra-luminous infrared galaxies (ULIRGs) at $z \sim 2$.  
Roughly 30\% of this IERO sample, 38 sources, are detected at X-ray, all of which are classified as AGNs  according to their X-ray luminosities in \cite{Xue:2011}. Among these X-ray IEROs, 79\% are dSFGs and the remaining 21\% are quiescents. We do note that quiescent galaxies with AGNs, which can also have strong 24 $\um$ emission \citep[see, e.g.,][]{Gu:2007}, may be misidentified as dSFGs. As shown in Figure~\ref{k36_ch15}b, several AGNs even show very blue K $-$ [3.6] contrary to their red [3.6] $-$ [24]. Further classification of these X-ray IEROs will be performed using their morphologies and rest-frame colors.

\section{Morphologies of IEROs from $HST$/WFC3 NIR Imaging}

Galaxy morphologies provide additional information on the mass assembly history of galaxies \citep{Bell:2005,Zhengx:2004}. 
%However, many previous morphological studies at $z \sim 1.5$ were constrained to blue galaxies (e.g., BM/BX and LBGs) and only probed rest-frame ultraviolet structures. With the new WFC3/IR observations, we can now assess rest-frame optical morphologies of a  large sample of red galaxies at $z \sim 2$ for the first time, permitting investigation of the true stellar structure of high redshift galaxies. 
We performed both visual classifications and  nonparametric morphological measurements to see whether quiescents and dSFGs in our sample can be separated based on their morphologies. All morphological analysis in this work are based on the CANDELS 6-epoch v2.0 drizzled mosaics for the WFC3 F160W band, with a pixel scale of $0.06''$, 
%a combination of both IR color and morphological criteria may yield a better classification.
\subsection{Visual Classification}
 \label{visual}
The IEROs in our sample show very  diversified morphologies, covering a wide range of  types from extended and disturbed low surface brightness features to bulge-dominated morphology.  We divide morphological types for this sample into three broad categories: 

{\bf \textit{Spheroid}}: Single, round, and centrally concentrated source with no evidence of extended low surface brightness features.

{\bf \textit{Disk}}: Undisturbed source with extended low surface brightness features.

{\bf \textit{Irregular/Merger}}:  Single highly irregular galaxy with evidence of non-axisymmetric, extended low surface-brightness features or two or more distinct galaxies showing distortions and interaction features such as tidal arms.

\begin{figure*}
\center{\includegraphics[width=0.9\textwidth]{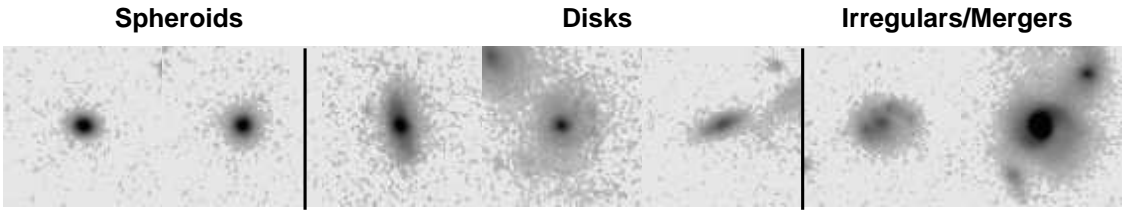}}%{paper1_mor_exam.eps}}
   \caption{ For the visual classifications, we simply divide the sample into three broad morphological types: spheroid on the left, disk morphologies in the middle, and irregular/mergers on the right panel. \label{mor_example}}
   \end{figure*}

Figure~\ref{mor_example} shows typical examples of the three types in the WFC3 F160W  images.
Visual inspections for this sample were performed by three of the co-authors independently. We combined the classifications from each inspector and reviewed the sources together to resolve the disagreements, which involved $\sim$ 15\% of the sample. The final visual morphological types are listed in Table~\ref{tab:mor}.
About 40\%  of this sample are classified as disk galaxies, 30\% are spheroids and the remaining 30\% are irregular/mergers. The X-ray sources
have the similar fractions of disk, spheroid and irregular/merger types: 45\%, 20\% and 35\% respectively. Most of the X-ray sources 
show a bulge component even they are identified as disk galaxies \citep[see also, ][]{kocevski:2011}. 

\begin{figure*}
 \center{\includegraphics[width=0.9\textwidth]{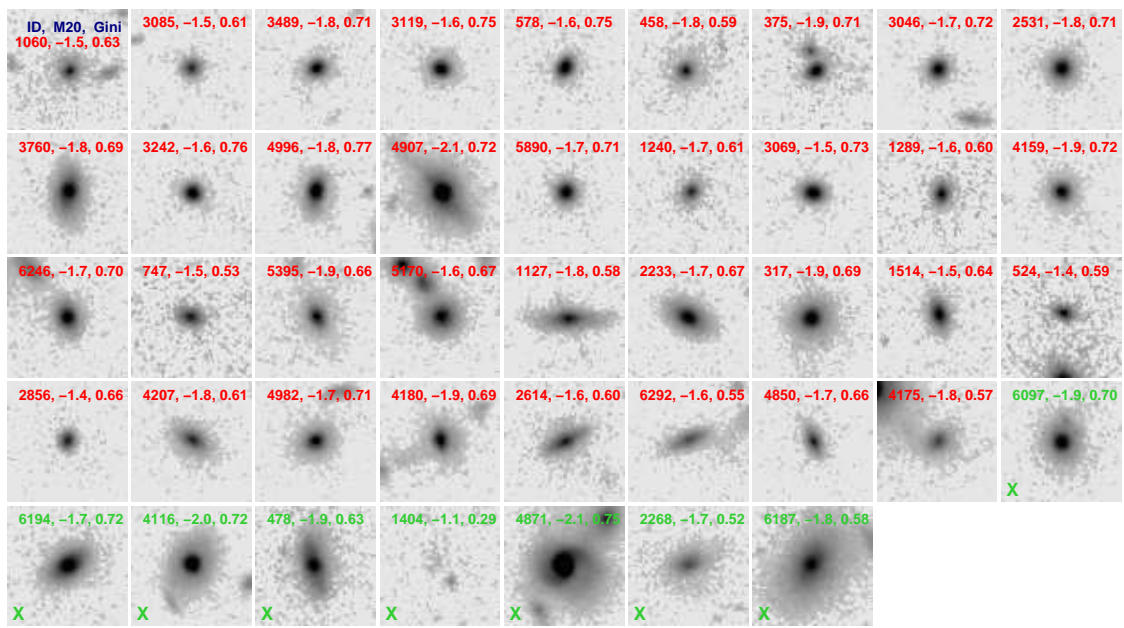}}%{paper1_h160_old.eps}}
   \caption{$HST$/WFC3 F160W images for the quiescent galaxies (quiescents) in our sample. Quiescents are defined by Eq 1 and lie within the dashed square in Figure~\ref{k36_ch15}. The source IDs (from FIREWORKS catalog), $M_{20}$, and Gini are labeled in each panel. X-ray sources are marked with an "X"  in the bottom-left corner. Both non-X-ray and X-ray sources are sorted in order of increasing $K - [3.6]$ colors. The size of each postage map is $4'' \times 4''$, and $1''$ corresponds to $\sim$ 8.5 kpc at $z \sim 2$. Nearly all objects have a prominent bulge component with $\sim$ 30\% of them  also showing an extended disk. The only outlier is Object 1404, which shows a very diffuse morphology with a photometric redshift $z \sim 3$. At this redshift, the MIPS 24 $\mu$m probes the rest-frame 6 $\mu$m, and even star-forming galaxies has low brightness. We thus argue Object 1404 is not a $z \sim 2$ quiescent galaxy.  
  \label{h_fire_eg}}
   \end{figure*}

\begin{figure*}

\center{\includegraphics[width=0.9\textwidth]{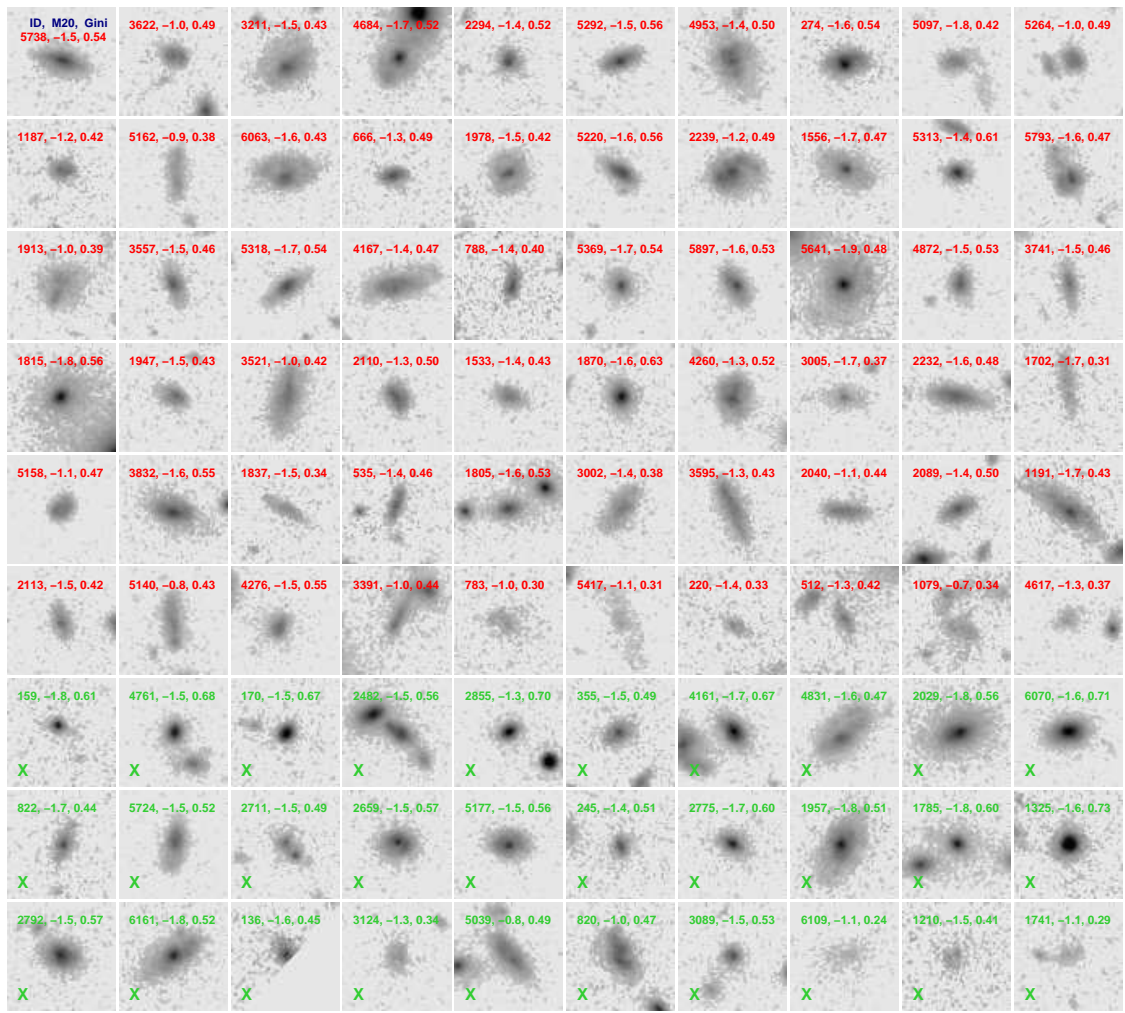}}%{paper1_h160_dusty.eps}}
\caption{ \textit{HST}/WFC3 F160W images for the dusty star-forming galaxies (dSFGs) in our sample. The dSFGs are all galaxies that lie outside the dashed square in Figure~\ref{k36_ch15}. X-ray sources are marked with  green "X". Fireworks ID, $M_{20}$, and Gini are shown in each postage stamp. Both non-X-ray and X-ray sources are sorted in order of increasing $K - [3.6]$ colors. Most of these dSFGs have disk or irregular/merger morphologies, with only a few of them, mainly X-ray sources, showing a strong bulge component.   \label{h_fire_sf}}
   \end{figure*}

We present F160W images for quiescents and dSFGs in Figures~\ref{h_fire_eg} and \ref{h_fire_sf}, respectively.  Roughly 70\% of the quiescent galaxies are spheroids, and 30\% of them show a significant disk component. The dSFGs, in contrast, are either regular disks or irregular/mergers, with only a few being spheroids. It should be noted that he quiescent disks are significantly different from the star-forming disks: the former also have a prominent bulge while the latter have small or no bulges. Thus we can separate the two populations based on whether they have a significant bulge component: Both spheroids and those disks with a prominent bulge are quiescents while the remaining galaxies are dSFGs.  
%In other words, all quiescents have a prominent bulge. 
%In fact, the most distinguishable morphological property between the two populations is whether they have a prominent bulge. We also derive their nonparametric morphological parameters for further classification.
\subsection{Nonparametric Morphological Parameters }
\label{quantitative}
Nonparametric morphological parameters, such as Gini and M$_{20}$, have now become popular in the automated identification of galaxy morphology. These parameters do not rely on certain model parameter fits, such as the S{\'e}rsic index, and therefore can be applied to irregulars as well as standard Hubble-type galaxies \citep{Lotz:2004}. 
The Gini coefficient is a concentration parameter, measuring the relative distribution of the galaxy pixel flux values. 
It is high for galaxies with much of their light concentrated in a small number of pixels, regardless whether those pixels are in the projected center. 
Therefore, the Gini method has the advantage that it do not require the galaxy to be circularly symmetric,
%Compared with the conventional measurement of galaxy concentration, concentration index $C$ \citep{Abraham:1994,Conselice:2003}, 
and is particularly useful for galaxies at high redshifts, which can have very distorted morphologies. $M_{20}$ describes the second-order moment of the brightest 20\% of the galaxy's flux, which is very sensitive to merger signatures such as multiple nuclei, tidal tails, and off-center star clusters. 
Gini and $M_{20}$ are shown to be closely related with visual morphologies for galaxies both at low and intermediate redshifts: early-type galaxies have higher Gini and lower M$_{20}$ while late-type galaxies have lower Gini and higher M$_{20}$ \citep{Lotz:2006,Abraham:2007,Capak:2007,Lotz:2008,Kong:2009}. At higher redshift, however, it is difficult to measure Gini and $M_{20}$ due to a lack of high-quality NIR (rest-frame optical) images. And their relationships with visual morphologies, as well as star-forming status remain unclear.
%Probing this relation at high-redshifts requires reliable measurements of Gini and M$_{20}$ at NIR (rest-frame optical) bands. for a representative sample of massive galaxies, which is now possible requies NIR (rest-frame) 
The new WFC3 data from CANDELS provides NIR imaging with sufficient angular resolution and depth, allowing us to reliably measure Gini and M$_{20}$ for a representative sample of massive galaxies at $z \sim 2$. 
%this relation holds, and how  parameters relation holds can still provide consistent results with visual classifications. especially for the  relationships remain unclear Several recent works extended the use of these parameters for galaxies of certain types at higher redshifts, e.g., LBGs, SMGs and DOGs \citep{Lotz:2007,Bussmann:2011,Swinbank:2010,Zamojski:2011}. With the new WFC3 data, it is now possible to we can  provides NIR (rest-frame optical) imaging with sufficient angular resolution and depth, allowing us to for a surficient large sample  
%but mostly restricted to certain types, e.g., LBGs, optically selected SFGs, and small numbers of SMGs and DOGs
%UV-bright samples or however, it was difficult to calculate these parameters for a representative sample of massive galaxies, most of which are optically faint \citep{van_dokkum:2006}. With the new NIR (rest-frame optical) data from CANDELS, such calculations become possible especially for the massive galaxies at $z \sim 2$ which are bright at NIR. 

\begin{figure*}
 
 \begin{center}
   \center{\includegraphics[width=0.9\textwidth]{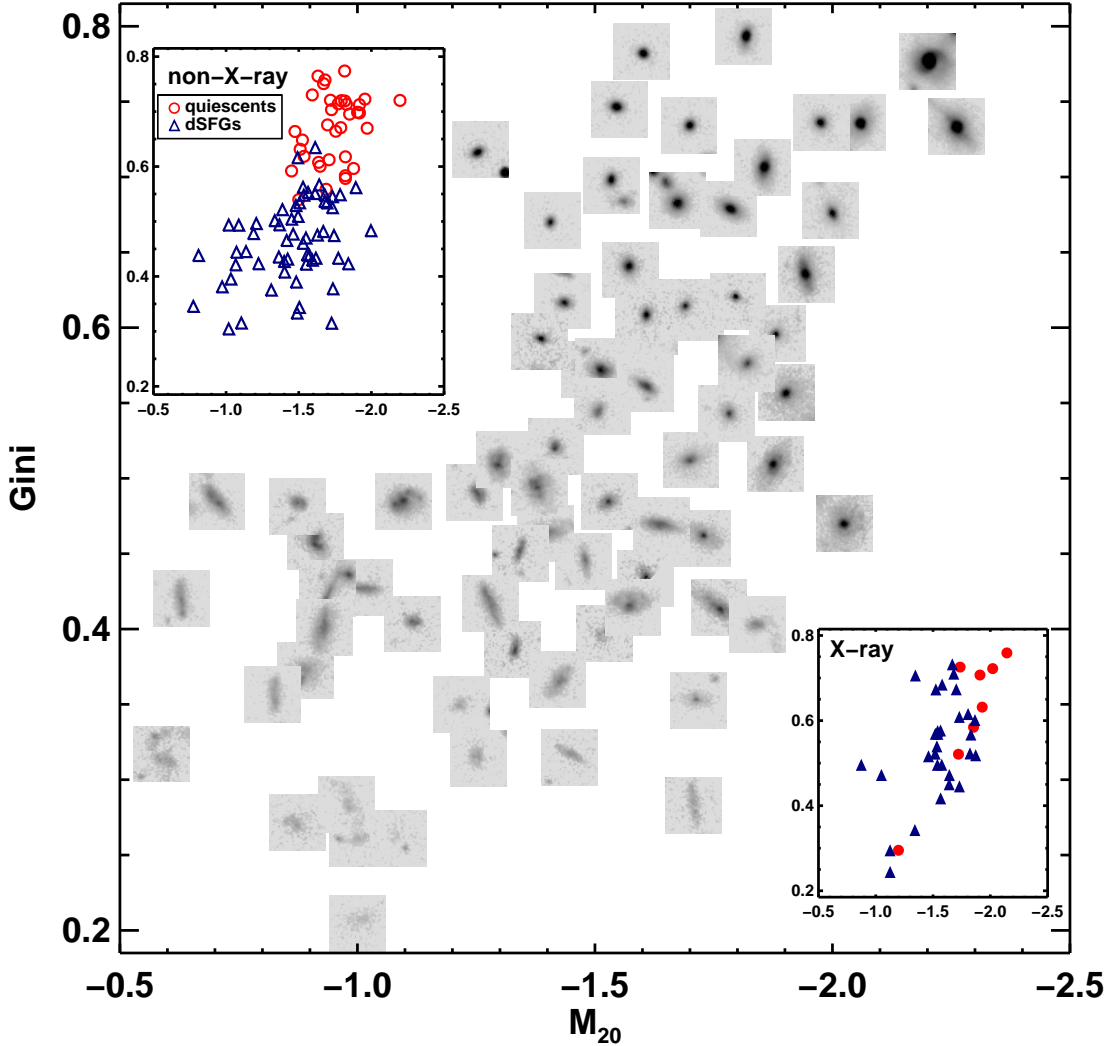}}%{paper1_gini_m20.eps}}
   \caption{Gini vs $M_{20}$ for the galaxies in our sample. Quiescents and dSFGs are defined based on Eq~\ref{eq:IR_color}. Non-X-ray sources are denoted with circles and triangles while X-ray sources  are denoted with filled circles and triangles, respectively. The quiescent and dSFGs are well separated  in this diagram (upper left insert). The postage-stamp images for a randomly selected subsample represent morphologies for galaxies in various location of this diagram. The quiescents have high Gini and low $M_{20}$, and the dSFGs  primarily have low Gini but cover a wide range of $M_{20}$. The quiescent and dusty star forming
galaxies with no X-ray detection are well separated by Gini = 0.58 with only a few outliers. The X-ray sources do not show such consistent classifications between morphologies and their IR colors.
  \label{G_vs_M20}}
   \end{center}
   \end{figure*}

We calculated Gini and M$_{20}$ for galaxies in our sample using the WFC3 F160W images following the definition of 
\begin{equation}
G = \frac{ \sum_i^N (2i -N -1) |f_i|}{(N-1) \sum_i^N |f_i|}
\end{equation}
where $N$ is the number of pixels within the images and $f_i$  are the fluxes for each pixel sorted in ascending order with $|f_1| \leq |f_2| \leq ... \leq |f_{N}|$, and 

\begin{eqnarray}
M_{20} \equiv {\rm log10}\left(\frac{\sum_j M_j}{M_{tot}}\right) & {\rm while } & \sum_j f_j <  0.2 \sum_j^N |f_j|
\end{eqnarray}
%~~~~and
\begin{equation} 
M_{tot} = \sum_j^N M_j = \sum_j^N f_j \cdot ((x_j - x_c)^2 + (y_j - y_c)^2)
\end{equation}
\citep{Lotz:2004,Abraham:2007} 
,where $x_c, y_c$ is the galaxy's center and $f_j$ are the fluxes for each pixel sorted in descending order with $|f_1| \geq |f_2| \geq ... \geq |f_{N}|$ \citep{Lotz:2004,Abraham:2007}.  Details of our procedure can be found in  \cite{Abraham:2007}. We shows the distribution of Gini and $M_{20}$ for our sample in Figure~\ref{G_vs_M20}, with randomly selected measurements marked by their stamp images.

\begin{figure*}
	\center{\includegraphics[width=0.9\textwidth]{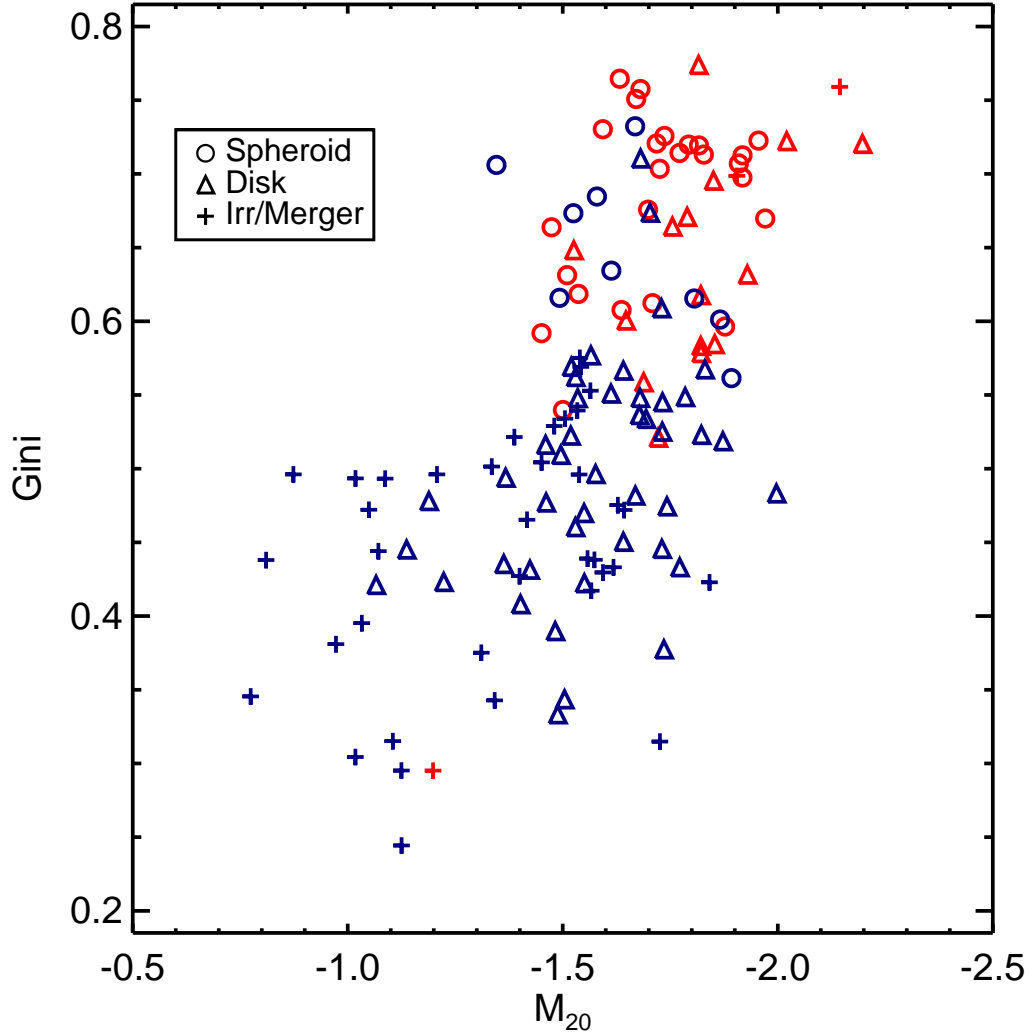}}%{paper1_gini_m20_mor1_small}}
	\caption{The distribution of all the IEROs with different visual morphological types in the $Gini-M_{20}$ diagram. Quiescents are denoted with red colors while dSFGs are in blue colors. Their distribution shows great agreement with the visual classification results. Spheroids have higher $Gini$ and lower $M_{20}$ while Irr/Mergers have lower $Gini$ and higher $M_{20}$. Interestingly, quiescent disks show primarily higher $Gini$ than the star-forming disks, consistent with that the former have a prominent bulge. 
	\label{visual_gini_m20}}
\end{figure*}

Galaxies of different visual morphological types occupy different regions in the Gini-M$_{20}$ diagram (Figures~\ref{G_vs_M20} and \ref{visual_gini_m20}). Spheroids have higher Gini and lower $M_{20}$, while most irr/mergers have lower Gini and high $M_{20}$. The disks span a wide range in both Gini and $M_{20}$: those disks with a prominent bulge tend to have higher Gini and lower $M_{20}$ whereas disks with small or no bulges have lower Gini and higher $M_{20}$. 

\begin{figure*}
  
  \center{\includegraphics[width=0.9\textwidth]{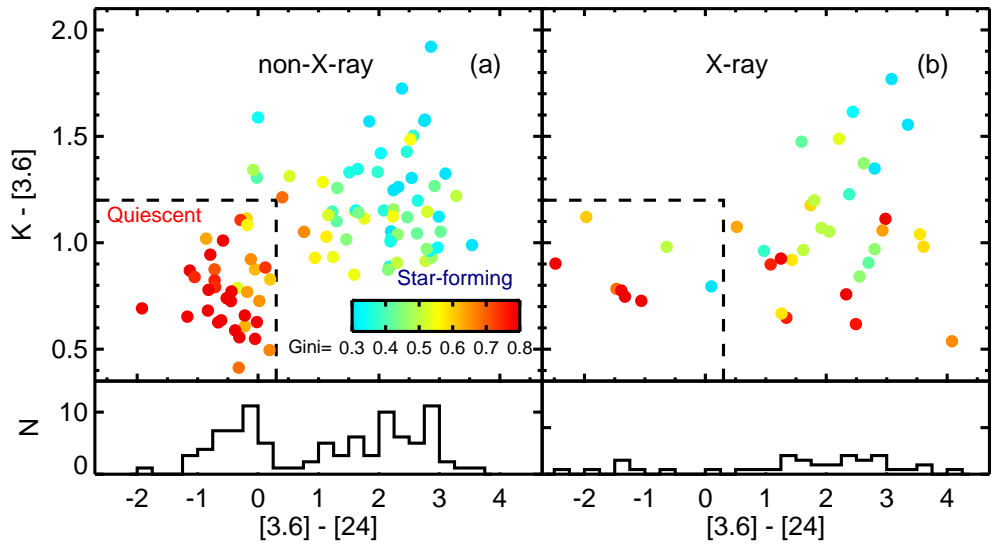}}%{paper1_ch15_k36_gini.eps}}
  \caption{ [3.6] $-$ [24] vs $K - [3.6]$ color-color diagram for the IERO sample with galaxies color-coded according to  the Gini coefficient. A nearly perfect correlation exists between [3.6] $-$ [24] color and Gini. Such a correlation, does not exist for the X-ray sources in the right panel.\label{k36_ch15_gini}}
 \end{figure*}

For the non-X-ray IEROs, quiescents and dSFGs are well separated in the Gini-M$_{20}$ diagram: nearly all quiescents have Gini $>$ 0.58, while dSFGs have Gini $<$ 0.58 with only a few outliers. To further verify this, we re-plot Figure~\ref{k36_ch15} but now with each object color-coded according 
to the Gini coefficient, as shown in  Figure~\ref{k36_ch15_gini}. This diagram shows that the IR color criteria used in Eq~\ref{eq:IR_color} to select quiescents also succeeded in choosing objects with Gini $\ga$ 0.58. The three sources with  [3.6] $-$ [24] $<$ 0.3 and red K $-$ [3.6] $>$ 1.2 have very low Gini coefficient, suggesting that they are not quiescents.  
% the tight correlation between the Gini coefficient and star forming activity for these massive galaxies indicates that the quenching of star formation must lead to or be accompanied by the increase of galaxy concentration. 

X-ray IEROs are plotted in Figure~\ref{k36_ch15_gini}(b). They show inconsistent classification between their IR colors and quantitive morphological parameters. While most of the X-ray sources in the quiescent region of Figure~\ref{k36_ch15} show higher Gini \footnote{The only one that shows very low Gini coefficient, ID 1404, is a high-z source, as discussed in Figure~\ref{h_fire_eg}.}, nine X-ray sources have Gini $>$ 0.58 yet have [3.6] $-$ [24]  $>$ 0.3, placing them outside the quiescent region. There are two possible reasons for these sources to cross the line: they are
quiescents with strong bulges and their 24 $\um$ emission is from central AGNs, or they are dSFGs with powerful AGNs appearing in their $H$-band images and causing an increase in Gini. We do notice that, however, most of these sources show a significant bulge except one source with ID 1325 appearing to be a point source (Figure~\ref{h_fire_sf}). In fact, among all the X-ray sources in our sample, 66\% (25 sources) of them have $L_{X} < $ 5 $\times$ 10$^{43}$ erg s$^{-1}$ while only 15\% (6 sources) have $L_{X} >$ 1 $\times$ 10$^{44}$ erg s$^{-1}$ \citep{Xue:2011}. Thus we argue that contamination from the AGN in optical bands is not significant for most X-ray sources. Those with high gini are likely caused by a significant bulge component. 
%Among We also find that there is no correlation between their X-ray luminosity and Gini coefficient.  
We will further verify this with their rest-frame colors in Section~\ref{rest}.

\begin{figure*}
	\center{\includegraphics[width=0.9\textwidth]{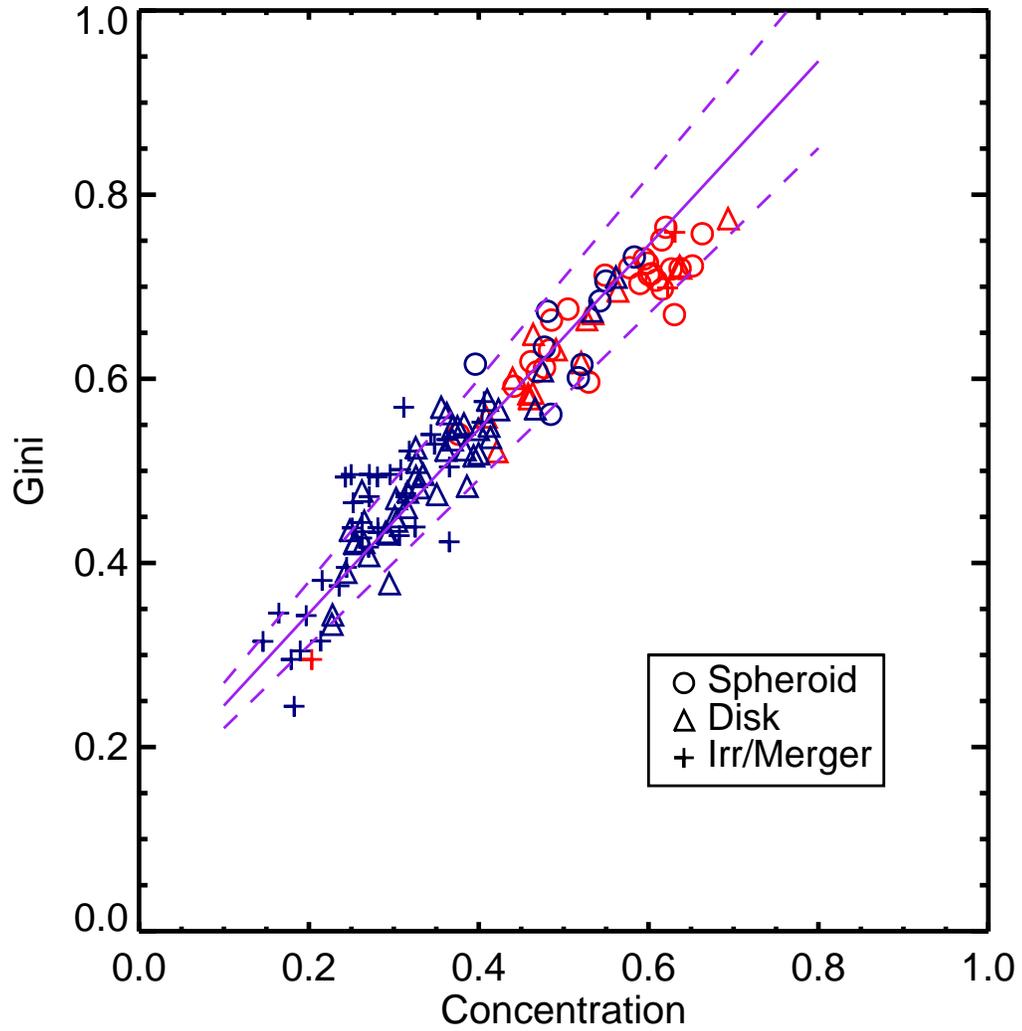}}%{paper1_gini_concentration}}
	\caption{Gini coefficient (Gini) vs. central concentration (C) for IEROs in our sample (points as in Figure~\ref{visual_gini_m20}). The
	solid line corresponds to unity slope, while the dashed lines are $\pm$ 10\% offset relative to the solid line. 
	\label{Gini_C}}
\end{figure*}

Though the adopted technique has been used and tested by many previous works \citep[see, e.g.,][]{Abraham:2007,Capak:2007,Kong:2009}, we evaluate the accuracy and reliability of Gini and $M_{20}$ in the following ways. We have already shown that there is good correspondence between visual morphological types and their distribution in the $Gini-M_{20}$ diagram (Figures~\ref{G_vs_M20} and \ref{visual_gini_m20}). Thus the $Gini-M_{20}$ system provides a reliable way to automatically distinguish different morphological types of galaxies in our sample. 
Many previous works show that below a certain S/N level $ <$ S/N $> \sim $ 2 , Gini and $M_{20}$ depend on the S/N ratio and is useless for distinguishing different populations \citep{Lotz:2004,Lisker:2008}. Benefiting from the bright cut, the median magnitude in the F160W band is $H \sim 22.5$. And all their images have a mean S/N per pixel $ <$ S/N $> >$ 2 with a median of $ < S/N > \sim 10$ (Lotz et al. 2012, in preparation). Therefore, the $Gini$ and $M_{20}$ for our sample do not suffer S/N effect. As an independent check, we also calculated galaxy central concentration, the concentration index ($C$, \citealt{Abraham:1994}), using also F160W images and compared it with $Gini$ (Figure~\ref{Gini_C}). Our calculated $C$ is closely correlated with $Gini$, consistent with that shown in previous works \citep{Abraham:2003}. This confirms that galaxies with high $Gini$ indeed are more concentrated. The $Gini-C$ relation shows larger scatter for objects with low concentration, mostly dSFGs. This is also expected: measurements of $C$ is based on simple aperture photometry, under the assumption that galaxies are circularly symmetric and have a well-defined center; however, neither of these assumptions can be fulfilled for the dSFGs in this sample. Thus we conclude that our calculated $Gini$ and $M_{20}$ provides reliable measurement of galaxy concentration and clumpiness for galaxies in our sample. 

%We also verified that by perturbing the $Gini$ and $M_{20}$ assuming a Gaussian distribution with

%We also verified that by perturbing our measured $Gini$ and $M_{20}$ using a Gaussian distribution by 10\% derived in this paper 

\section{Photometric Redshifts and Stellar Masses}
\label{photoz}

\subsection{Photometric Redshifts}
\label{sed_fitting}
Only $\sim $ 23\%, 31 galaxies, of the IEROs in our sample have spectroscopic redshifts ($z_{spec}$), and therefore we derive photometric redshifts ($z_{ph}$) for the rest using the standard SED fitting technique. After separating our IEROs into quiescents and dSFGs, we are able to select SEDs accordingly to do SED fitting.

%and we use this knowledge as a prior to determine their redshifts: as shown below, for the two populations we use different SED types according to their spectra types.
  
\begin{figure*}
 \center{\includegraphics[scale=0.9]{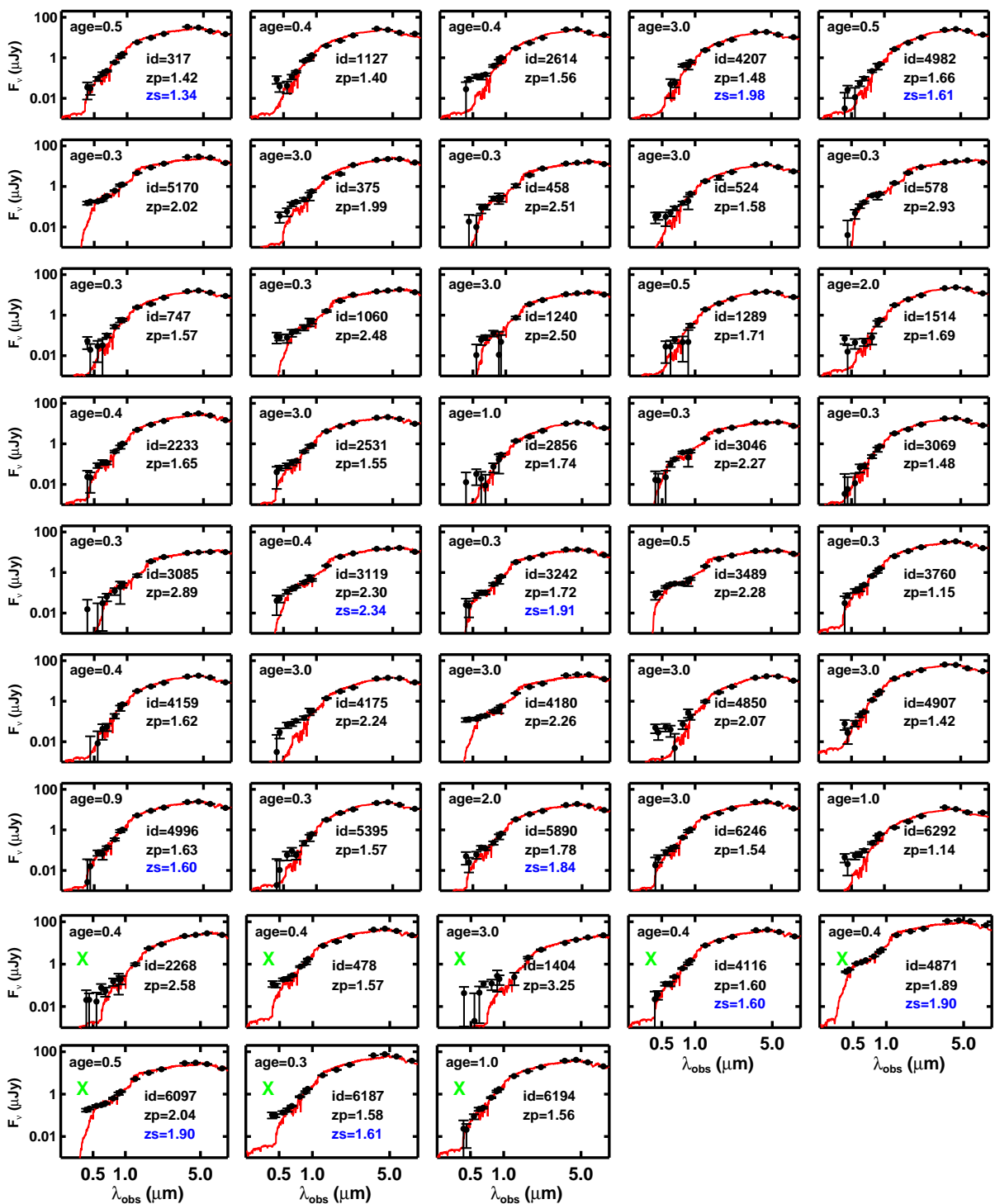}}%{paper1_photoz_eg.eps}}
   \caption{SED fitting to measure photometric redshifts for the quiescents with  SSP models as templates. The source ID, age of the best-fit SSP model , photometric redshift ($z_{ph}$) and spectroscopic redshifts($z_{spec}$, if available) are labeled in each panel. Most of these quiescents can be well fit with SSP model templates with ages ranging from 0.4 Gyr to the age of the Universe of 4 Gyr at that redshift. "X"s denote the X-ray sources in Figure~\ref{h_fire_eg}. 
   \label{eg_fit}}
   \end{figure*}

We used the single stellar population (SSP) models of \citet[hereafter BC03]{bc03} to construct a template set for the quiescents. We assumed solar metallicity and Calzetti extinction law \citep{Calzetti:2000} in the range of  $0\leq A_V\leq 0.5$~mag.  For the 10 quiescents with spectroscopic redshifts, we obtain very good agreement between $z_{ph}$ and $z_{spec}$, with the normalized  median absolute deviation  $\sigma_{NMAD}$\footnote{Defined as  1.48 $\times$ median($|z_{spec} - z_{ph}|/(1+ z _{spec})$) \citep{Brammer:2008}} =  0.03.
Figure~\ref{eg_fit} shows that the SEDs of the quiescent IEROs in our sample are well fit by these SSP models, with best-fit ages ranging from 0.4 Gyr to 4 Gyr. Many of them also show a rest-frame UV excess, which is likely due to an additional young stellar population component \citep[see also][]{Yanh:2004}. However, the UV excess does not affect the resulting photometric redshifts, which are mainly determined by the rest-frame optical-to-MIR SEDs.

\begin{figure*}
\center{\includegraphics[scale=0.9]{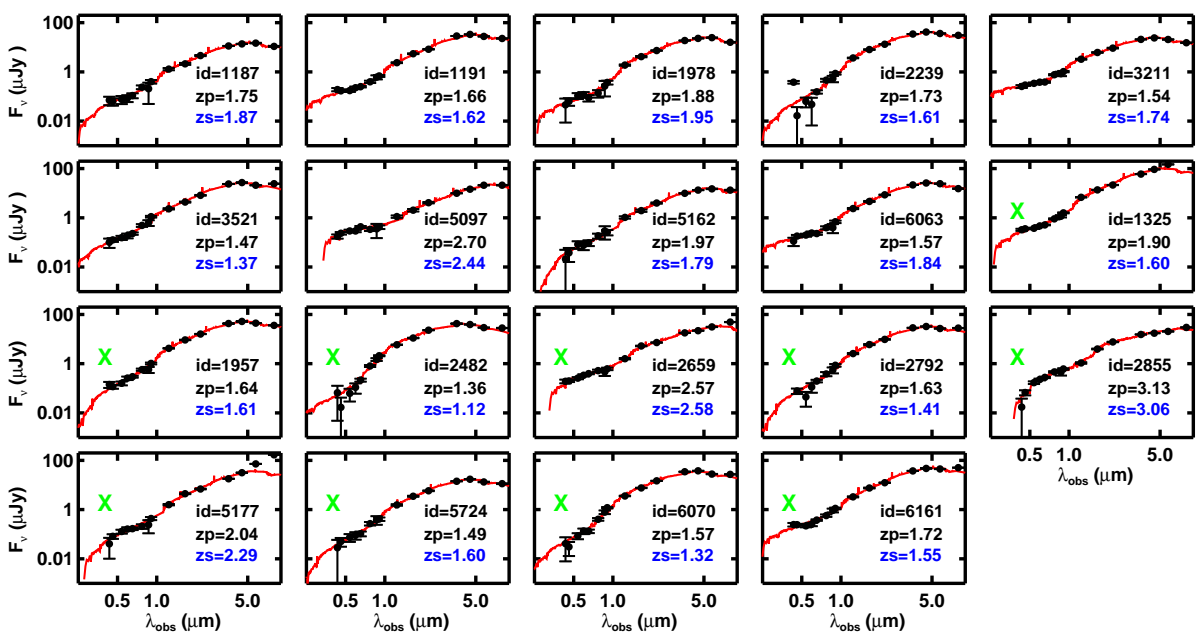}}%{paper1_photoz_sf.eps}}
\caption{Example SEDs  for the dSFGs with EAzY spectral fits overlaid. Here we only show the SEDs for those dSFGs with spectroscopic redshifts. Data points  with 1$\sigma$ errors are shown in black. The red line shows the linear combination of template SEDs that best fit the data. The source ID, $z_{ph}$ and $z_{spec}$ are labeled in each panel. 
\label{sf_fit}}
\end{figure*}

The dSFGs are  much more diverse in that they span a wide range in their $[3.6] - [24]$ colors and show a large variety of morphological characteristics. We fit these dusty galaxies using the  "Easy and Accurate Zphot from Yale" \citep[EAzY;][]{Brammer:2008}. We allow for a linear combination of six templates  which can recover the entire rest-frame color space occupied by observed galaxies. An accuracy of  $\sigma_{NMAD} = 0.07$ was obtained for the 19 dSFGs with spectroscopic redshifts, and their SEDs are shown in Figure~\ref{sf_fit}.

%We do not choose to use particular templates  for the X-ray sources. We have verified that it did not produce better redshift estimates either by including additional AGN templates in the template set of EaZY or by using pure AGN templates for the X-ray sources in this sample. It is mainly because that AGNs  do not contribute significantly in the optical bands for most of these sources, as discussed in Section~\ref{quantitative}. Moreover, by applying the template error function, the longer wavelength data (e.g., 5.8 and 8.0 $\um$), which are more likely to  suffer contamination from AGNs, have lower weight in the SED-fitting.
%The default template error function is adopted to account for wavelength-dependent template mismatch. By applying this template error function, different wavelength regions are given different weights: the optical bands are assigned higher weights while the UV and NIR, where the SED fitting is most uncertain, have lower weights.

\subsection{ Stellar Masses}
\label{stellarmass}
To obtain stellar masses, we fit the observed B-to-4.5 $\micron$ SEDs of both the quiescent and dSFGs using "Fitting and Assessment of Synthetic Templates" \citep[FAST;][]{Kriek:2009a}. We fixed their redshifts at the derived redshift, and fit the SEDs to a set of simple stellar population models. These models have exponentially declining star-formation histories (SFH) with $0.1 < \tau < 10$ Gyr and $0 \leq A_{V} \leq 4$ assuming Calzetti law. We adopted a 
Salpeter \citep{Salpeter:1955} initial mass function and solar metallicity. The stellar masses derived from the best-fitting template are then scaled to a Kroupa IMF \citep{Kroupa:2001} by dividing the Salpeter stellar mass by a factor of 1.6 \citep{Marchesini:2009}.
%Salpeter \citep{Salpeter:1955} initial mass function and solar metallicity.
% We then scaled the stellar masses derived from the best-fitting template to a Kroupa IMF \citep{Kroupa:2001} by dividing the Salpeter stellar mass by a factor of 1.6. 
We have verified that by using the \cite{Maraston:2005} model instead of BC03 models, the stellar mass is systematically smaller with an offset of $-0.20$ and a scatter of 0.17. This is consistent with previous studies for $z \sim 2$ samples \citep{Muzzin:2009,Marchesini:2009}. We also compared our stellar mass estimates with that derived using the approach in \cite{Wuyts:2007,Wuyts:2008}, which fitted three different star formation histories: a single stellar population (SSP) without dust, a constant star formation (CSF) history with dust ($A_{v}$ varying from 0 to 4), and an exponentially declining star formation history with an e-folding timescale of 300 Myr and the same range of $A_{v}$. The two methods give fairly consistent results with a offset of $-0.02$ and a scatter of 0.13.

\begin{figure}
\center{\includegraphics{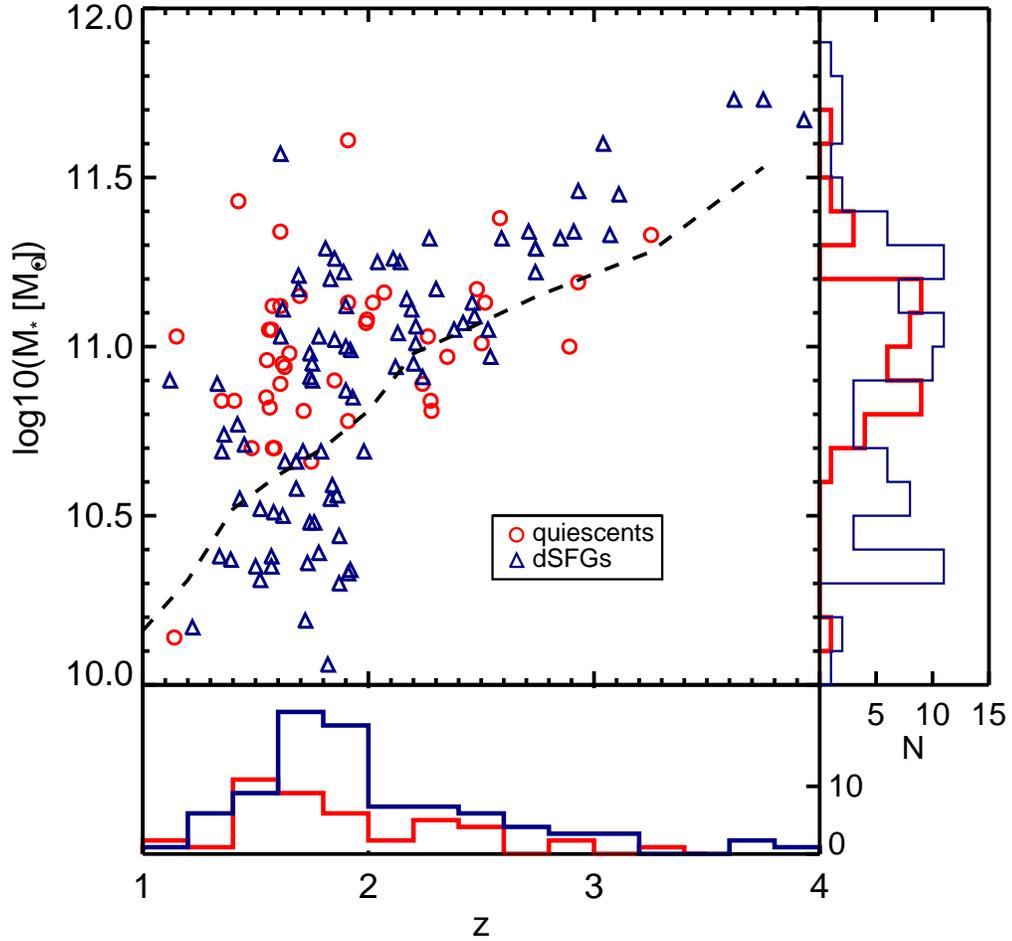}}%{paper1_lmass_z}}
\caption{Redshift and stellar mass histograms of the full  sample. The dashed line corresponds to the mass completeness limit for a 3.6 $\um$ selected sample ($[3.6] < 21.5$) at $1 < z < 4$. The redshift range of the IERO sample spans from 1 to 4 with most of them located at $1.5 < z < 2.5$, indicating that the simple color criterion of $z_{850} - [3.6] > 3.25$ can effectively select massive galaxies at $z \sim 2$. 
\label{lmass_z}}
\end{figure}

The mass and redshift distributions are shown in  Figure~\ref{lmass_z}. We also derive the mass completeness limit as a function of redshift for a 3.6 $\um$ selected ($[3.6] < 21.5$) sample. This completeness limit is calculated using the full and deeper FIREWORKS catalog following the method in \cite{Marchesini:2009}. 
While these IEROs span a wide range of
$1\leq z_{ph} \leq 4$, most of them are in $1.5 < z < 2.5$ within a narrow peak centered at
$z_{ph}=1.8$. Their derived stellar masses are more striking, with a median value of
$\sim 1\times 10^{11} M_{\odot}$. All this is in broad agreement with the prototype
IERO sample of \cite{Yanh:2004}. These IEROs are clearly among the most massive
galaxies. At $M > 10^{11} M_{\odot}$ (where the our sample is nearly complete; see
below), the fraction of quiescents increases significantly from $\sim 22$\% at 
$2\leq z\leq 2.5$ to $\sim 43$\% at $1.5\leq z\leq 2$ (Table 1), suggesting
that we are seeing the quenching process in action in the most massive systems \citep[also see, e.g.,][]{Cameron:2011}.

\begin{figure}
\center{\includegraphics{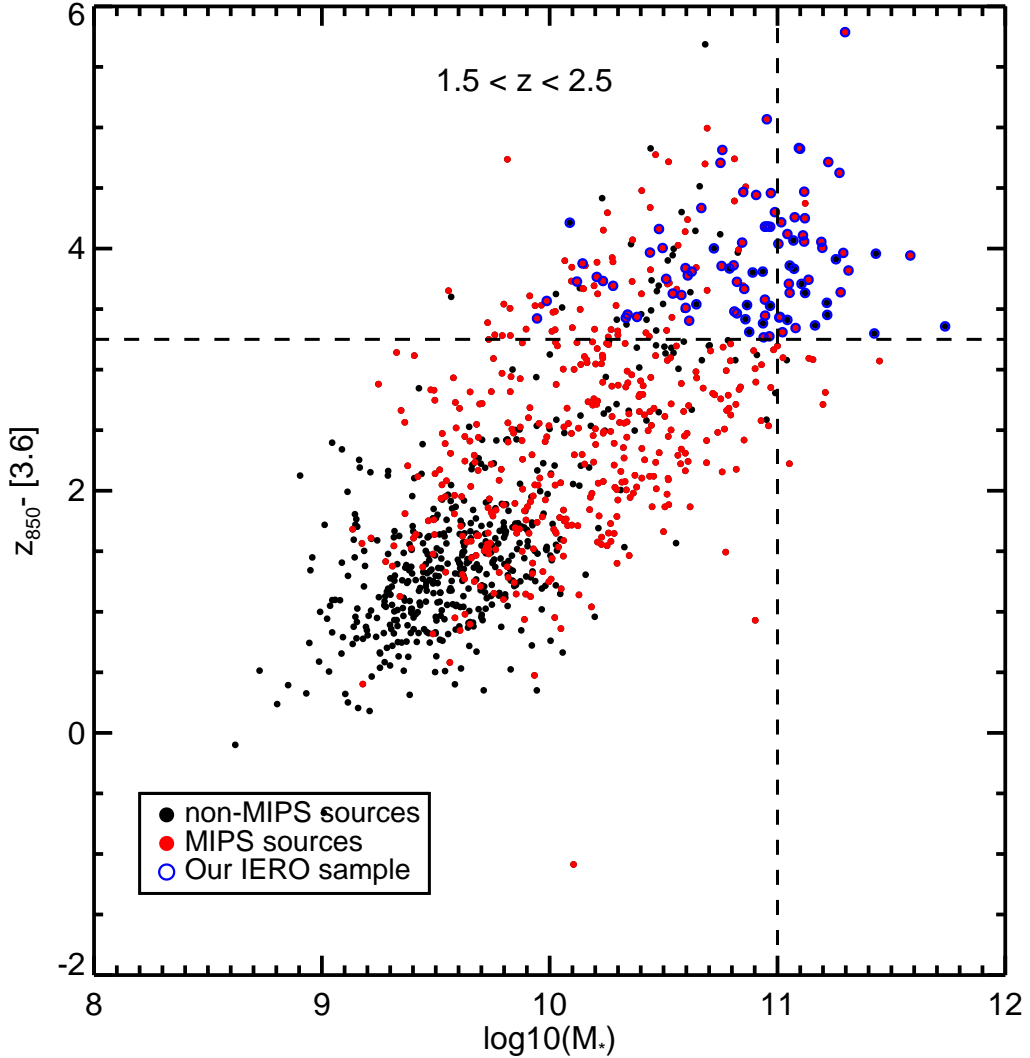}}%{paper1_mass_limit.eps}}
\caption{$z_{850} - [3.6]$ vs stellar mass for all galaxies at $1.5 < z < 2.5$ in the FIREWORKS catalog. Red points are MIPS sources. 
The purple open squares are IEROs in our sample ($z_{850} - [3.6] > 3.25$ and $[3.6] < 21.5$). There are 49  galaxies with $M_{*} > 1 \times 10^{11} \Msol$, which all have $[3.6] < 21.5$. About 80\% of  them (39) are selected as IEROs with  $z_{850} - [3.6] > 3.25$. Among the nine galaxies not selected as IEROs, four of them are X-ray sources \citep{Luo:2008,Xue:2011}, which also have the highest MIPS 24 $\um$ flux and bluest $z_{850} - [3.6]$ colors: $z_{850} - [3.6] < 3.1$. Their SEDs have strong AGN components which may cause their stellar mass to be overestimated. The remaining five galaxies just missed being IEROs according to $z_{850} - [3.6]$ being too blue. Only one of them is not detected at  24 $\um$. 
        \label{mass_limit}}
  \end{figure}

It is important to understand the completeness of this sample in terms of stellar mass. For this purpose, we compare our work to that of FIREWORKS \citep{Wuyts:2008}, where the K-band selected sample is complete
at $M\geq 10^{10.5} M_{\odot}$ to $z \sim 3$. The mass completeness of a flux-limited sample is
clearly a function of redshift, and here we concentrate on the redshift range of 
$1.5\leq z\leq 2.5$ because this is where we are most interested. As stated above, the approach for deriving stellar mass for the FIREWORKS galaxies gives consistent results as our estimates. 
We find that our IERO sample has recovered $\sim$ 80\% of the FIREWORKS galaxies at 
$M\geq 10^{11} M_{\odot}$ within $1.5\leq z\leq 2.5$, as shown in  Figure~\ref{mass_limit}. 

The other 20\% (nine galaxies) have $z_{850} - [3.6] < 3.25$~mag and thus are missed by our
selection. Among these nine galaxies, four of them are X-ray sources (Luo et
al. 2008; Xue et al. 2011), and they also have the highest MIPS 24~$\um$
fluxes. The SEDs of these four objects have strong AGN components, which may have
caused their stellar masses to be overestimated. The other five
galaxies have just slightly bluer $z_{850} - [3.6]$ colors than the IERO
criterion, and only one is not detected in MIPS 24~$\mu$m. They are clearly at
the border of being selected as IEROs. We hence argue that the IERO selection
using $z_{850} - [3.6] > 3.25$~mag and $[3.6] < 21.5$~mag is nearly complete 
at $M_{*}\geq 10^{11} M_{\odot}$ at $1.5\leq z\leq 2.5$ except for the bluest galaxies.

\section{Rest-frame Properties of IEROs}
\label{rest}

%joint distributions of .....

\begin{figure*}
  \center{\includegraphics[width=0.9\textwidth]{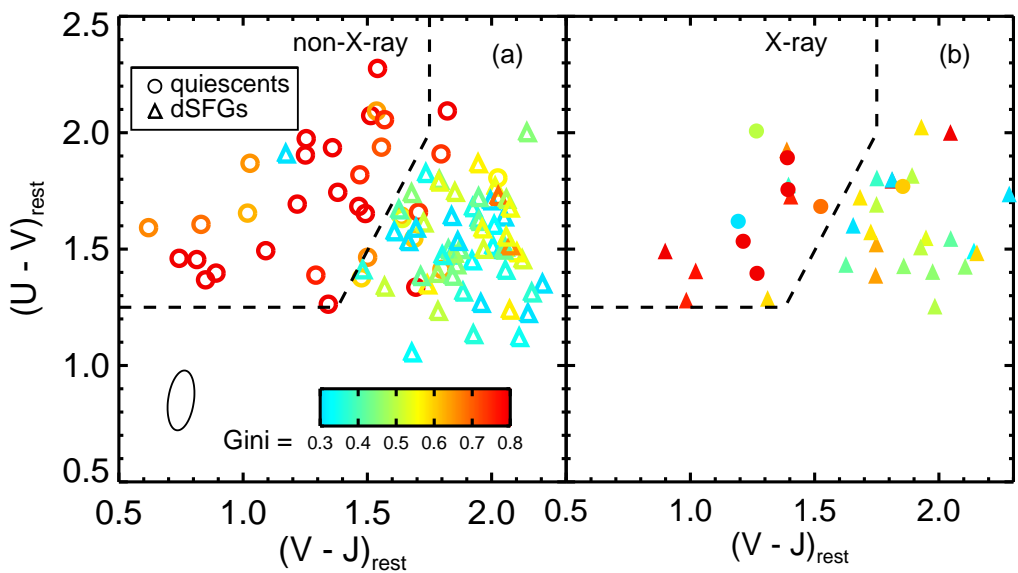}}%{paper1_uv_vj_gini.eps}}
   \caption{Rest-frame $U - V$ vs. $V - J$ color-color diagram for non-X-ray (left) and X-ray (right) IEROs in our sample. The error ellipse is shown in the lower left of the left papel. Objects are color-coded based on
Gini coefficient. Quiescent and dSFGs can be well separated in this diagram with the dashed line. The locations of high- vs. low-Gini objects on opposite side to the lines confirms that morphology and rest-frame color-color yield generally consistent classifications for both populations. The distributions  overlap somewhat, but mostly near the boundary. Both  massive quiescent and dusty galaxies have red $U - V$ colors, but the quiescents are bluer in $V - J$, implying they are not suffering significant dust extinction. X-ray sources with different Gini values are about as well separated as non-X-ray sources (right panel).\label{uv_vj_gini}}
  \end{figure*}

The rest-frame colors permit additional characterization of galaxies. \cite{Wuyts:2007} and \cite{Williams:2009} proposed that quiescents and dSFGs can be separated in rest-frame $U - V$ versus $V - J$: while both quiescents and dSFGs show red $U - V$ colors either due to old stellar populations or dust extinction, quiescents are bluer in $V - J$ due to lack of dust reddening. We computed these two colors by interpolating observed data following the algorithm  of \cite{Rudnick:2003}, as shown in Figure~\ref{uv_vj_gini}. The rest-frame colors uncertainties were calculated using the input photometry through Monte Carlo simulations. For each galaxy, we created mock SEDs using a Gaussian distribution and calculated their rest-frame $UVJ$ colors. This procedure was repeated 1000 times to derive the uncertainties for their color estimation. The average 1$\sigma$ error ellipse is shown in the lower left of Figure~\ref{uv_vj_gini}a.
 
Quiescents and dSFGs in the sample can  be well separated in the $UVJ$ diagram (Figure~\ref{uv_vj_gini}a), consistent with our classification based on the observed IR colors and morphological parameters. However, there are a few quiescents lying just across the dividing line which may be due to either photometric errors or that they suffering extinctions. While we can locate these galaxies in the quiescent region by pushing the dividing line to redder $V - J$ colors, many dSFGs will also enter the quiescent region. In fact, we verified that the observed $K - [3.6]$ colors for these outliers are also redder than other quiescent galaxies at similar redshifts. Interestingly, most of these outliers show edge-on disk morphologies in their F160w images, e.g., ID 1127, ID 2614, and ID 6292 (Figure~\ref{h_fire_eg}), indicating that they suffer more severe dust extinction. 
X-ray sources can also be separated in the $UVJ$ diagram (Figure~\ref{uv_vj_gini}b): those X-ray sources with high $Gini$ are in the quiescent regions, despite that several of them are classified as dSFGs based on the IR colors. We argue that these X-ray sources with high $Gini$ may have significant bulges, yet their 24 $\um$ mainly comes from AGN.% egde on disks 

\begin{figure*}

  \center{\includegraphics[width=0.9\textwidth]{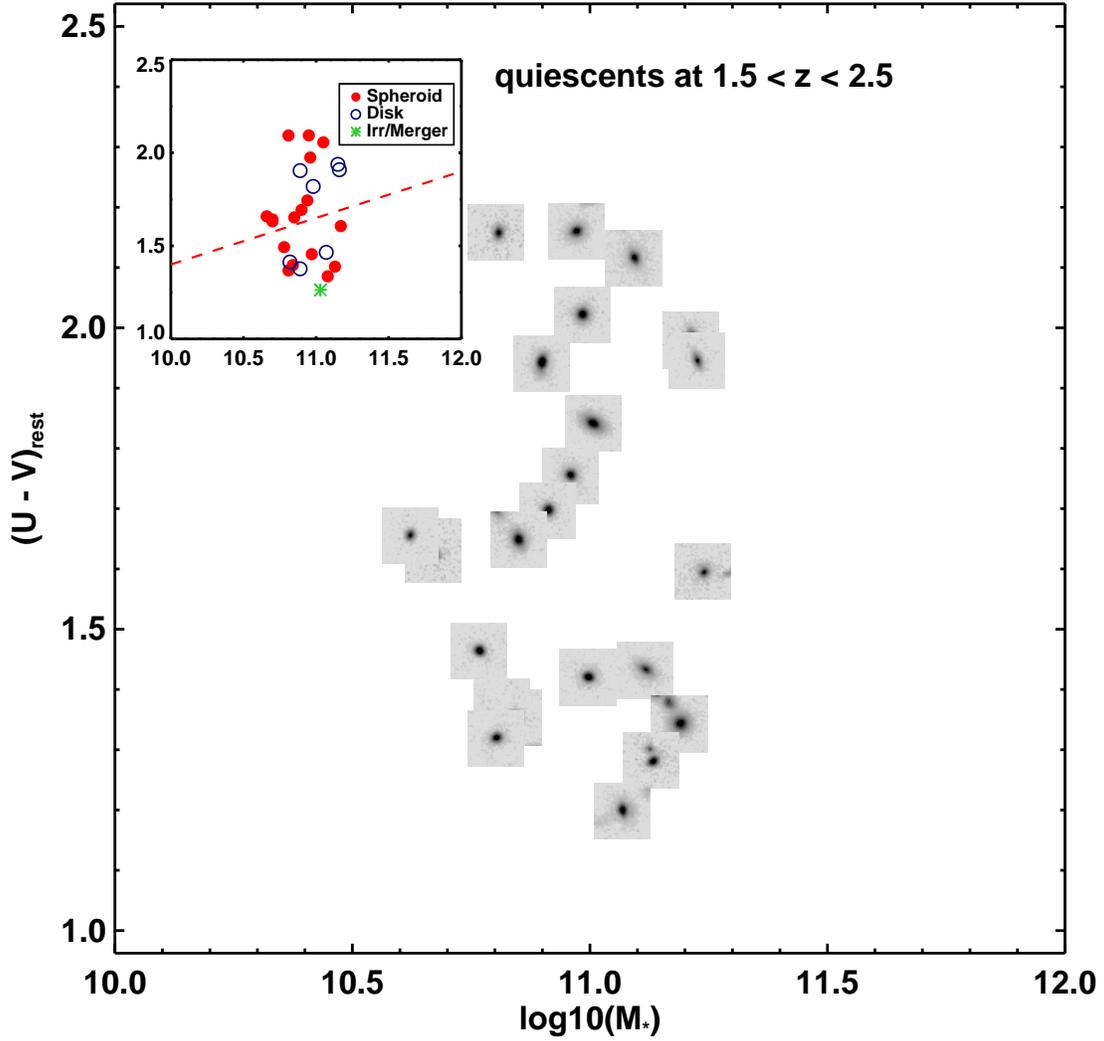}}%{paper1_mor_uv_lmass_eg_noagn.eps}}
   \caption{\small Rest-frame $U - V$ color vs stellar mass  for  quiescents in $1.5 < z < 2.5$ in the sample.  \textit{HST}/WFC3 F160W images  are also shown. Visual classification results for these quiescents are shown in the upper left panel. The dashed red line  is the red sequence criterion at $z$ = 2 extrapolated from the same criterion at $z < 1$ in \cite{Borch:2006}. \label{uv_lmass_eg}}
  \end{figure*}

\begin{figure*}

   \center{\includegraphics[width=0.9\textwidth]{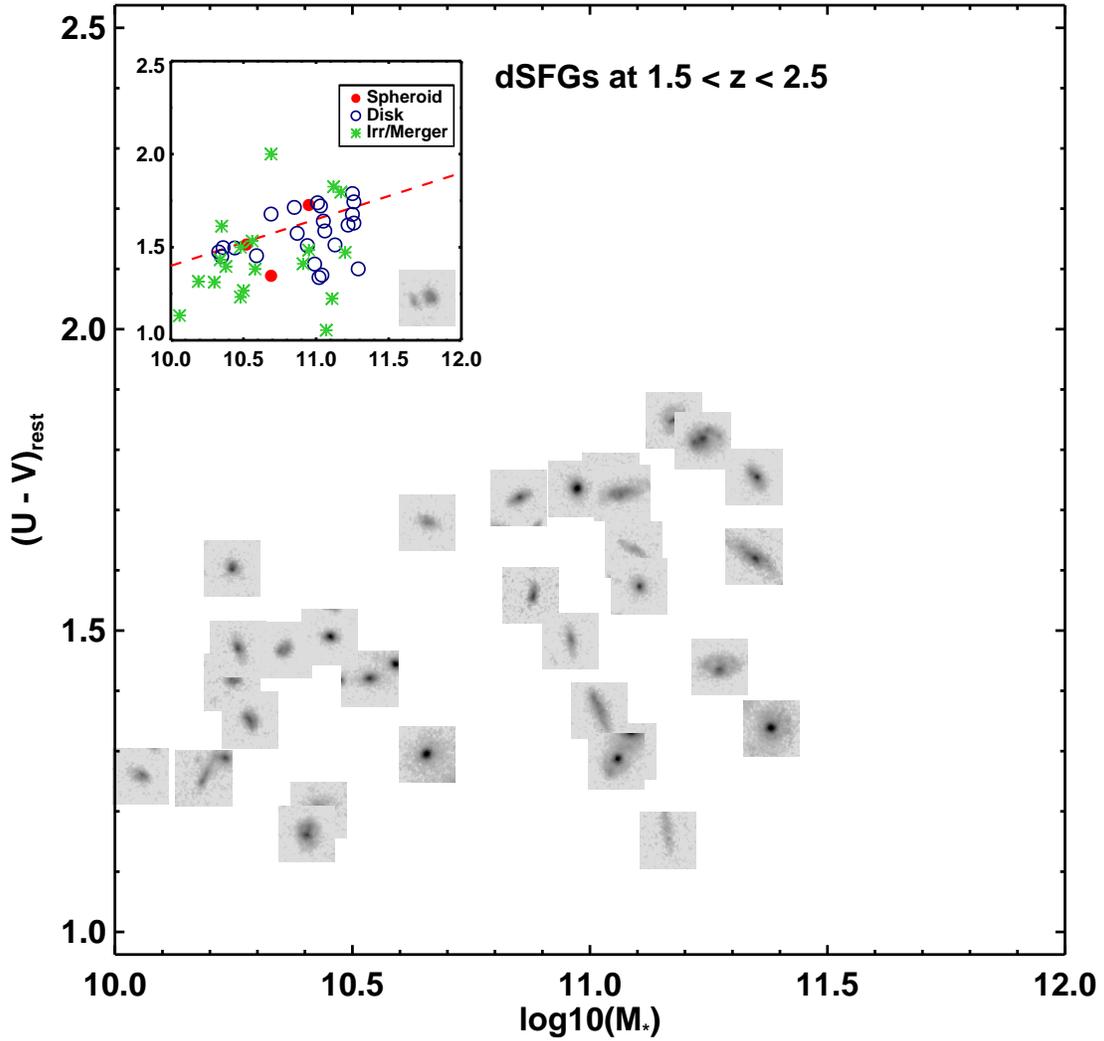}}%{paper1_mor_uv_lmass_sf_noagn.eps}}
   \caption{\small  Rest-frame $U - V$ color vs stellar mass  for  dSFGs at in $1.5 < z < 2.5$ in the sample, also shown are \textit{HST}/WFC3 F160W images for a randomly selected dSFGs sample. Visual classification results for the dSFGs are shown in the upper left panel.  Most of the dSFGs with $M_{*} > 1 \times 10^{11}\Msol$ show  disk morphologies. 
      \label{uv_lmass_sf}}
  \end{figure*}

Quiescents and dSFGs can not be separated in the color-mass diagram (CMD) as shown in Figure~\ref{uv_lmass_eg} and Figure~\ref{uv_lmass_sf}. We confine our analysis on their distribution in the CMD to the non-X-ray IEROs in $1.5 < z < 2.5$, where our sample is nearly complete for the most massive galaxies.
The dashed red line is the red sequence criterion at z = 2 extrapolated from the same criterion at z < 1 in \cite{Borch:2006}. Both quiescents and dSFGs in this sample are massive and red. Particularly, Figure~\ref{uv_lmass_eg} shows that the $U -V$ colors span  a wide range for quiescents. It is also interesting to note that the distribution of spheroids and disks appears similar in the CMD. These quiescent disks contribute $\sim$ 30\% of the whole quiescent population in $1.5 < z < 2.5$, confirming pervious findings \citep{van_der_Wel:2011,Cameron:2011,Kartaltepe:2011}. %All the quiescent disks also show a significant bulge, different from those star-forming galaxies.
%Another important fact emerging from Figure~\ref{uv_lmass_eg} is that there seems no difference in the distribution of 
%quiescents with or without a disk component in the color-mass diagram. The formation mechanism of these massive quiescent disks  need further investigation. 
%It thus indicates that the age of their stellar populations may be quite different, consistent with the large age spread found in Sec~\ref{sed_fitting} \citep[also see, e.g.,][]{Whitaker:2010}. 
%Another important fact emerging from Figure~\ref{uv_lmass_eg} is that
%A 2-dimensional Kolmogorov-Smirnov test gives a probability of P = 0.23 that the two kinds of galaxies are drawn from the same sample.
For the dSFGs, we find that disk galaxies tend to be more massive and have redder $U - V$ than the Irr/Mergers (Figure~\ref{uv_lmass_sf}). 
%The probability that these two kinds of galaxies are drawn from the same sample is only P = 0.1, which is derived from a 2-dimensional Kolmogorov-Smirnov test. 
In fact, as shown in Table~\ref{tab:mor}, 70\% of the massive dSFGs are disks. These disks are very extended and do not show clear signatures of mergers/interactions from visual inspection. $M_{20}$ for most of these disks is smaller than that for Irr/Mergers (Figure~\ref{G_vs_M20}), suggesting that they are relatively undisturbed. These properties are incompatible with the expected compact or highly perturbed morphologies of ongoing mergers. Instead, the stochastic accretion of gas may play a greater role in the formation of massive disks \citep{Forster:2006,Dekel:2009,Martig:2009,Genzel:2006,Genzel:2011,Forster:2011b}.

\section{Conclusions}

%present new insights except confirming 

With the new $Chandra$ 4 Ms imaging and $HST$/WFC3 F160W imaging from CANDELS, we studied relationships between SEDs, morphologies, and star-formation status for a sample of 133 IRAC-selected Extremely Red Objects (IEROs) with $z_{850} - [3.6] > 3.25$~mag and $[3.6] < 21.5$~mag in the GOODS-S. 
%We identify a sample of 133 IRAC-selected Extremely Red Objects (IEROs) in GOODS-S using $z_{850} - [3.6] > 3.25$~mag and $[3.6] < 21.5$~mag. 
These criteria
select high-mass galaxies at $z \ga 1.5$, and our sample is nearly mass-complete at 
$M_{*}>10^{11} M_\odot$ within $1.5\leq z\leq 2.5$. The IEROs show a clear bimodality in their IR colors, and we find that they can be classified into quiescent galaxies and dusty star-forming galaxies according to their star-formation activities. Such a classification produces very consistent results with that using $Gini$ and rest-frame colors ($U - V$ vs. $V - J$), which indicates that these quantities are closely related. 
Roughly 67\% of IEROs in this sample are dusty star-forming galaxies while the remaining 33\% are quiescent galaxies. At $M > 10^{11} M_{\odot}$, the fraction of quiescents increases significantly from $\sim 22$\% at $2\leq z\leq 2.5$ to $\sim 43$\% at $1.5\leq z\leq 2$.
Among the whole IERO sample $\sim $ 30\%, or 38 sources, are AGNs, and the identification of these AGNs are shown to be very important in understanding the discrepancies among different classification methods. A number of galaxies are identified as star-forming galaxies based on their 24 $\um$ emission, yet are in the quiescent region of the $UVJ$ diagram; we show that most of them are actually AGNs. 
%The inclusion of X-ray data is shown to be very important to identify quiescent galaxies yet with AGNs, which would otherwise be selected as dSFGs based on IR colors. the discrepancies  
%In particular, we are the first to perform a detailed analysis of both visual qualitative and non-parametric quantitative rest-frame optical morphologies of a well-defined and large sample of $z \sim 2$ massive galaxies. 
%We show unambiguous evidence that there is a strong correlation between morphology and star-formation status for these  massive galaxies.% suggesting a "Hubble sequence" is already in place since $z \sim 2.5$.
%This result nicely complements recent  morphological work relying on Sersic fitting \citep[i.e., a model-dependentapproach,][]{Kriek:2009b,Wuyts:2011b,Bell:2011}, and suggests that a "Hubble sequence" is already in place by $z \sim 2.5$.

% correlation between galaxy morphology and the position in the mass-SFR also exists. While overall the dSFGs in our sample follow the "main sequence" of star-formation, the Irr/Mergers tend to be less massive and  above the main sequence while the disks tend to be massive and below the main sequence. This suggest that the star formation excess in the high-SSFR outliers may be largely due to galaxy interactions/mergers. The relatively low SSFR in the massive disks implies that quenching of star formation may first happen in these massive galaxies, i.e., "downsizing".

The morphological study in this work provides important clues to massive galaxy evolution at $z \sim 2$. The quiescent and dSFGs in this sample show distinctive morphologies. While all quiescents have a prominent bulge, most of the dSFGs are either disks or irregular/mergers with no or small bulges. This suggests that a prominent bulge is necessary to form a massive quiescent galaxy. We also show that the two populations can be well separated based on the Gini coefficient. The quiescents have higher Gini coefficients than the dSFGs, indicating that the formers are more concentrated. We argue that the quenching process for star formation must lead to or is accompanied by the increase of galaxy concentration.  

Roughly 30\% of the quiescents also show an extended disk component, confirming recent findings with a larger sample. Moreover, we also find that $\sim$ 70\% of massive dSFGs with $M_{*} > 10^{11} M_\odot$ at $1.5 \leq z\leq 2.5$ are classified as disk galaxies. Most of these disks appear relatively undisturbed with no clear signatures of mergers/interactions, based on both visual inspection and quantitative measurements. The prevalence of disks among massive galaxies at $z \sim 2$ challenges the merging scenario as the main mode of massive galaxy formation, and suggests that the stochastic accretion of gas plays a greater role.

This work is based on observations taken by the CANDELS Multi-Cycle Treasury Program with the NASA/ESA HST, which is operated by the Association of Universities for Research in Astronomy, Inc., under NASA contract NAS5-26555. We gratefully acknowledge R. Abraham for access to his morphology analysis code. This work is also supported under the National Natural Science Foundation of China under grants (10878010, 10221001, 10873012 and 10633040) and the National Basic Research Program (973 programe no. 2007CB815404 and 2007CB815405).

%\bibliographystyle{apj}
%\bibliography{iero}

%\begin{thebibliography}{111}
%\expandafter\ifx\csname natexlab\endcsname\relax\def\natexlab#1{#1}\fi

 %These spheroidal X-ray  dSFGs tend to have bluer $K - [3.6]$ colors and higher Gini, implying that they may suffer less dust reddening. 

%\begin{figure}
%\center{\includegraphics{fig10}}%{paper1_zp_zs}}
%\caption{Spectroscopic redshifts vs. photometric redshifts for the sample. The photometric redshifts for the two populations are derived  using templates that are appropriate to their star-forming class (quiescent vs. dSFGs respectively). The normalized standard deviation of photometric redshifts for quiescent and star-forming galaxies  are  0.03 and 0.07 respectively (see text for definition of $\sigma_{NMAD}$). 
%\label{zs_zp}}
%\end{figure}

 %\begin{figure}
 % \center{\includegraphics{fig13}}%{paper1_mass_sfr_morp.eps}}
 %  \caption{Galaxy morphologies in the SFR-mass diagram for IEROs at $1.5 < z < 2.5$. We include only non-X-ray sources to exclude sources that may have significant contribution from AGN to the IR luminosity. Quiescent galaxies and dSFGs are denoted with filled circles and triangles, respectively. Red, blue, and green color correspond to  $Spheroid$, $Disk$ and $Irr/merges$ according to our visual classification. The dashed line is the average SFR-mass relation (main sequence, MS) for a sample of massive galaxies at $z \sim 2$ from GOODS.  A  correlation between galaxy morphology and  star-formation activity is clearly seen in this diagram. We also find that those dSFGs above the MS (high-SSFR outliers, or starbursts) tend to be $Irr/Mergers$, indicating that they have a different star formation mode, and such a (starburst) mode can be largely attributed to interaction/mergers.  \label{mass_sfr}}
 % \end{figure}

      \begin{deluxetable}{lllll}
%\tabletypesize{\footnotesize}%small}%scriptsize}
\tablecolumns{5}
\tablewidth{0pt}
 \tablecaption{Visual classification results for IEROs in our sample.\label{tab:mor}}
\tablehead{ \colhead{Type} & \colhead{All} & \colhead{Spheroids} & \colhead{Disks} &
  \colhead{Irr/Mergers}}

\startdata
Total           &  133 (38)\,\tablenotemark{a}  & 34 (8) & 59 (17) & 40 (13) \\ 
Quiescents & 43 (8) & 25 (2) & 15 (4)   & 3   (2) \\
dSFGs            & 90 (30)& 9   (6)    & 44 (13) & 37 (11)\\
Massive\,\tablenotemark{b} Quiescents at 1.5 $\leq z <$ 2.0  & 10 (6) & 4 (2)  & 5 (3)& 1 (1) \\
Massive Quiescents at 2.0 $\leq z <$  2.5 & 4 (0)& 2 & 1 & 1 \\
Massive dSFGs        at 1.5 $\leq z <$  2.0 & 13 (5) & 1 (1) & 8 (3) & 4 (1)\\
Massive dSFGs        at 2.0 $\leq z <$ 2.5    & 14 & 0   & 11 & 3 (1)\\

\enddata

\tablenotetext{a}{The number of X-ray sources.}
\tablenotetext{b}{$M_{*} > 10^{11} M_{\odot}$, where our sample is nearly complete.}
\end{deluxetable}

\end{document}